\documentclass[aps, reprint, superscriptaddress, longbibliography]{revtex4-1}
\usepackage{latexsym,amssymb,amsfonts,mathrsfs,amsmath,bm,ulem}
\usepackage{graphicx}
\usepackage[usenames,dvipsnames]{xcolor}
\usepackage{soul}
\usepackage[linktocpage,colorlinks=true,linkcolor=blue,citecolor=blue,breaklinks=true,urlcolor=blue]{hyperref}

\renewcommand{\vec}[1]{\bm{#1}}

\AtBeginDocument{
    \newwrite\bibnotes
    \def\bibnotesext{Notes.bib}
    \immediate\openout\bibnotes=\jobname\bibnotesext
    \immediate\write\bibnotes{@CONTROL{%
    apsrev41Control,author="08",editor="1",pages="1",title="0",year="1"}}
     \if@filesw
     \immediate\write\@auxout{\string\citation{apsrev41Control}}%
    \fi
}
\begin{document}

\title{Engineering reconfigurable flow patterns via surface-driven light-controlled active matter} 

\author{Xingting Gong}
\affiliation{Department of Applied Physics, Stanford University, 348 Via Pueblo, Stanford, CA 94305, USA}

\author{Arnold J. T. M. Mathijssen}
\affiliation{Department of Bioengineering, Stanford University, 443 Via Ortega, Stanford, CA 94305, USA}

\author{Zev Bryant}
\affiliation{Department of Bioengineering, Stanford University, 443 Via Ortega, Stanford, CA 94305, USA}

\author{Manu Prakash}
\email[Correspondence: ]{manup@stanford.edu}
\affiliation{Department of Bioengineering, Stanford University, 443 Via Ortega, Stanford, CA 94305, USA}

\date{\today}

\begin{abstract}
Surface-driven flows are ubiquitous in nature, from subcellular cytoplasmic streaming to organ-scale ciliary arrays. Here, we model how confined geometries can be used to engineer complex hydrodynamic patterns driven by activity prescribed solely on the boundary. Specifically, we simulate light-controlled surface-driven active matter, probing the emergent properties of a suspension of active colloids that can bind and unbind pre-patterned surfaces of a closed microchamber, together creating an active carpet. The attached colloids generate large scale flows that in turn can advect detached particles towards the walls. Switching the particle velocities with light, we program the active suspension and demonstrate a rich design space of flow patterns characterised by topological defects. We derive the possible mode structures and use this theory to optimise different microfluidic functions including hydrodynamic compartmentalisation and chaotic mixing. Our results pave the way towards designing and controlling surface-driven active fluids.
\end{abstract}

\maketitle

\section*{Introduction}

The ability of biological organisms to self-assemble and organize has inspired new ideas in engineering and physics. Unlike traditional condensed matter systems at equilibrium, a defining characteristic of active matter is the injection of energy at the local scale of its constituents, which then cascades upward to give rise to emergent phenomena at larger scales \cite{sriram_review,marchetti_review,activenematics_review,needleman_review, vicsek1995,flocks_scalefree,toner2005hydrodynamics,sliusarenko2007aggregation, wu2009periodic, shaevitz2015, 2012bsub_turbulence, mathijssen2019collective}. The self-organizing capability of active matter makes it a fertile ground for new design principles and technologies. However, the realization of such devices is currently contingent upon our limited ability to control or program active materials \cite{woodhouse2017active,woodhouse2012spontaneous, goldstein2008microfluidics, ross2018}.

As a strategy for designing tunable active matter, we are inspired by the prevalence of surface-driven activity in nature. Rather than programming activity in the bulk, one could potentially prescribe what is on the boundary which in turn modulates and control bulk flows. For example, in human airways the coordinated motion of micron-scale cilia across the entire organ drives coherent flows and is essential for mucus clearance \cite{ciliaclearance, elgeti2013metachronal}. Furthermore, cytoplasmic streaming in the Characean algae is a salient example of surface-driven activity at the subcellular scale -- organelle-carrying myosin motors walk along fixed actin tracks, resulting in macroscopic circulation of the cytoplasm \cite{goldstein_review,goldstein2008microfluidics,woodhouse2013cytoplasmic}. Synthetic examples of active surfaces include artificial cilia \cite{den2008artificial, van2009printed}, self-propelled droplets and colloids accumulated on walls \cite{bricard2015emergent, bechinger2016active, zottl2016emergent, maass2016swimming}, Quincke rollers \cite{quincke}, engineered bacterial carpets \cite{darnton2004moving, mathijssen2019nutrient, jin2018biofilm}, and molecular motility assays \cite{schaller2010polar}. 

The ability to micro-manipulate flow structures is of great interest for applications such as lab-on-a-chip devices \cite{dittrich2006lab, schneider2011algorithm}. 
Yet, miniaturising self-contained microfluidic devices that do not require external macroscopic pumps and valves has remained a major challenge in the field. A solution could be to instead generate flows internally, without conventional pumps, by injecting momentum from patterned active surfaces \cite{mathijssen2019nutrient}, but little is understood how such topological patterns affect the flow properties and structures that can emerge across the scales.
In other words, while external pressure driven  microfluidics are now ubiquitous in research and industrial applications, the design space of internally driven flows using surface activity in confined geometries is almost entirely unknown.
Here we present a simple set of activity patterns of surface defects for patterning complex spatial and temporal surface activity. Simple defect geometry and dynamics enables a rich set of complex flow structures inside the bulk fluid that can be programmed by boundary effects, leading to possible new microfluidic functions in confined geometries. 

Inspired by recent advances in optogenetics, we consider patterning surface activity with light. Light has in recent years proven to be a powerful means of achieving spatiotemporal control living systems and study of active matter due to its ability to target behavior of active components with high spatiotemporal resolution \cite{ross2018}. 
Of particular interest to us, engineered cytoskeletal motors incorporating a photosensitive LOV2 domain have been shown to modulate their speed or direction in response to blue light \cite{nakamura2014remote, ruijgrok2020newpaper}. These motors have also been used for spatiotemporal control of active liquid crystals \cite{zhang2019structuring}.

In the first half of the paper, we develop an in-silico design tool to explore light-controlled active surface-driven flows in confined geometries.
Specifically, we consider active particles that walk along filament tracks fixed to the boundary of a flow chamber, where the direction of motion along the filament is switchable upon illumination. Viscous drag on the active particle leads to momentum transfer into the bulk, creating macroscopic flows characteristic of cytoplasmic streaming, and further redistributing the particles \cite{goldstein_review,oocyte_streaming_review,monteith2016mechanism, hydrogels_cytoskeleton,dogic2012}. We first analyze the simplest emergent flow structure generated by the motion of many such particles in a flow chamber. We then perturb the system with light (such as reversal of velocity) to explore the design space of possible flow structures. In the second half of the paper, we introduce a analytical framework based on an interior squirmer model to further generalize these flow structures.  We demonstrate that surface-driven activity can achieve remarkably complex and time varying 3D flow structures with properties like chaotic mixing and compartmentalised particle confinement with no physical barriers. Overall, our results provide insight into boundary-driven flows in naturally occurring biological systems and pave the way for using surface activity defects to program re-configurable bulk flows at small scales.

\section*{Methods}
\begin{figure*}
\includegraphics[width=\textwidth]{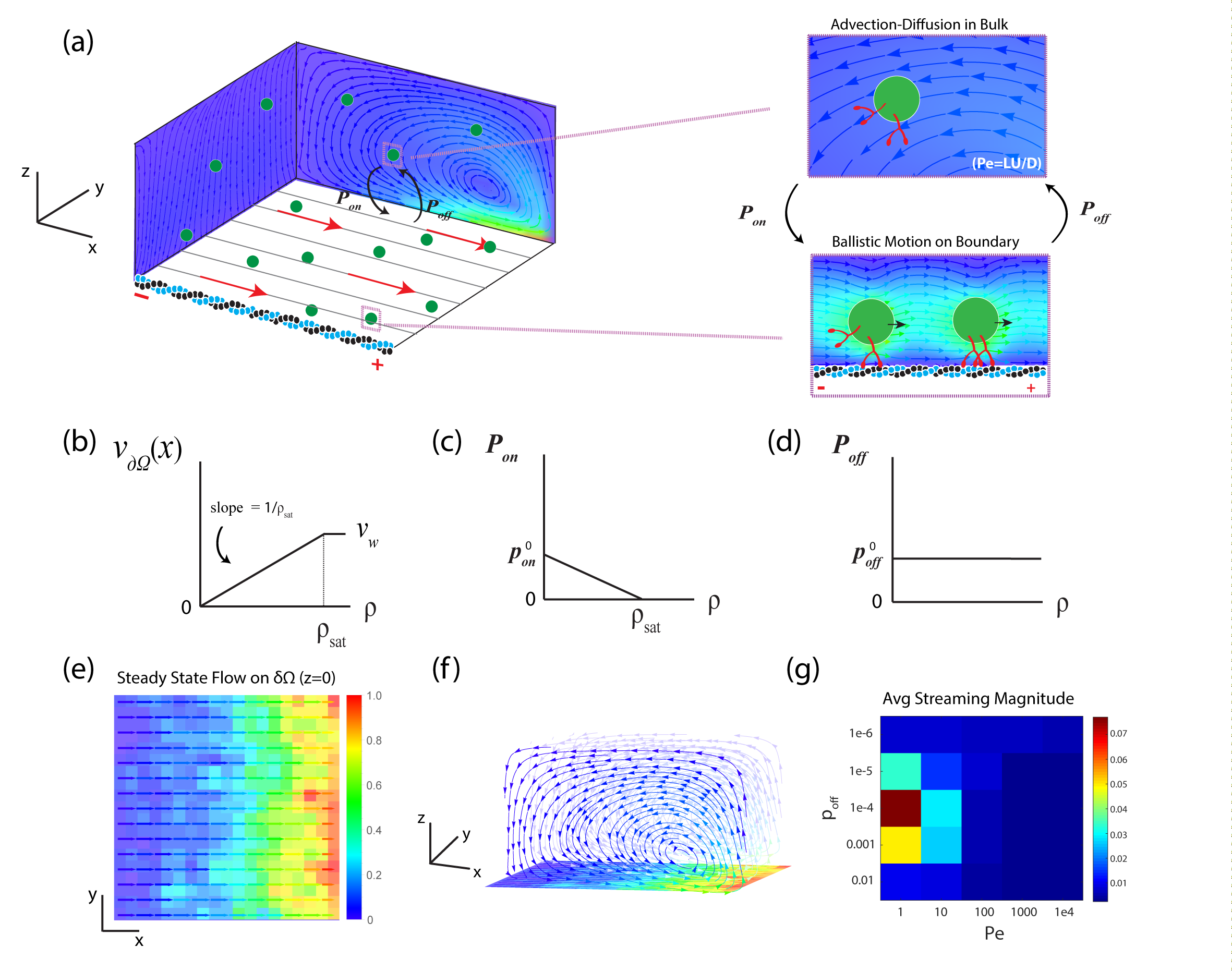}
\caption{\label{Fig1} 
Model for surface-driven flows in confined volumes. 
\textbf{(a)}
A confined chamber of size $N_x \times N_y \times N_z = 20 \times 20 \times 10$ that consists of a single active surface at $z=0$, uniformly coated with parallel and polarized tracks. The gray lines denote the orientation of particle trajectories on the surface (i.e. orientation of actin filaments patterned on the surface). Active colloidal particles (green) bind and unbind from the surface with rates $P_\text{on}$ and $P_{\text{off}}$. Bound particles walk ballistically in a fixed direction along the tracks, imparting momentum on the surrounding fluid and creating macroscopic flows. Unbound particles are free to advect and diffuse in the bulk, with relative strength set by the P\'eclet number $Pe$. 
\textbf{(b)} The velocity at each grid point $v_{\partial \Omega}(x)$ on the boundary increases linearly with particle concentration and saturates to the particle walking speed $v_w$. 
\textbf{(c)} The attachment rate $P_{\text{on}}$ decreases linearly to zero at the saturating concentration $\rho_{\text{sat}}$, a value which is set by the grid and particle size, while 
\textbf{(d)} the detachment rate $P_{\text{off}}$ is modeled as a constant. 
\textbf{(e-g)} Simulation results for a simple case: uniform flow in a confined chamber. All results are from shapshots of simulations recorded after simulating for $2 \times 10^6$ time steps to approach steady state, with each time step $dt \approx 10^{-6} L/v_w$ sec.
\textbf{(e)} The magnitude of the velocity at each point on the boundary for a specific choice of parameters (the optimum in Fig. 1G) shows an accumulation of particles toward the right edge of the chamber. The color bar depicts the flow magnitude, scaled by $v_w$ to a maximum of one.
\textbf{(f)} The existence of a wall forces the fluid in that region upward, creating a 2D vortex in the $xz$-plane. 
\textbf{(g)} Phase diagram of $P_{\text{off}}$ and $Pe$, showing that high streaming velocities in a confined volume favors low $Pe$ and an intermediate detachment rate. The average streaming velocity for each set of parameters is calculated across the full chamber. $P_\text{off}$ is reported in units of probability per time step (see SI for a more detailed description of parameters).
}
\end{figure*}

In this work, we present a simulation framework to explore flow-patterns generated by active surface driven flows. We set up the simulation by first considering a rectangular chamber, discretised using a grid of $N_x \times N_y \times N_z$ cells, where a single surface is patterned by tracks that are parallel and polarized in orientation (e.g. actin filaments) [Fig.~\ref{Fig1}A]. Colloids coated with active particles akin to molecular motors ([Fig.~\ref{Fig1}A green particles) are suspended in bulk and can attach to the surface, after which they walk ballistically from the minus to the plus end of the tracks in the absence of light, and reverse direction in the presence of light. As these particles move along the surface, they impart forces on the fluid and thus generate long-ranged circulating flows, also referred to as streaming. Particles in the bulk are advected by these flows which, in turn, can transport them towards the surface, where they bind and unbind from the surface through probabilities of attachment or detachment.

To solve for the system dynamics, we alternate between (1) computing the flow field $\bm{u}(\bm{r})$ at position $\bm{r}$ using a CFD solver, (2) integrating the bulk particle motion with Brownian dynamics, and (3) updating the surface particle density and dynamics [see Supplementary Information for details]. For the first step, inspired by Lighthill and Blake \cite{lighthill1952squirming, blake1971spherical}, we implement a slip velocity on the active surface due to the motion of the bound particles [Fig.~\ref{Fig1}B]. 
In the dense particle limit, the surrounding flow will saturate to the particle walking speed, but in the sparse limit the velocity vanishes. In between, we ignore inter-particle interactions and assume that the flow magnitude is approximately linear with respect to particle concentration [see Suppl. Fig.~S1].

Detached particles in the bulk are subject to advection and diffusion. Their relative strengths are set by the P\'eclet number $Pe =LU/D$, where $U$ is the characteristic velocity of the motors (e.g. $v_w$ in Fig.~\ref{Fig1}B), $L$ is the longest chamber length and $D$ is the diffusion constant. We solve for particle trajectories in the bulk by integrating the Langevin equation in the overdamped limit:
\begin{equation}
r_i(t+dt) = u_i(t)dt +  \sqrt{2Ddt}~\eta_i(t),
\end{equation}
where $i$ refers to components of the position $\vec{r} = \{x,y,z\}$ in Cartesian coordinates, the time step is $dt$ and $\vec{\eta}(t)$ is uncorrelated Gaussian white noise defined by $\langle \eta_i(t) \rangle=0$ and $\langle \eta_i(t) \eta_j(t') \rangle=\delta_{ij}\delta(t-t')$ in terms of the Kronecker and Dirac delta functions.

Particles that approach the surface closer than a distance $\epsilon$ can attach with rate $P_{\text{on}}$, modelled as a linearly decreasing function of particle density [Fig.~\ref{Fig1}C]. 
We assume that the particles are otherwise non-interacting. On the boundary, particles walk ballistically with a direction specified by the tracks (such as a surface coated with filaments). Conversely, bound particles can detach from the surface with a constant rate $P_{\text{off}}$  [Fig.~\ref{Fig1}D]. Initially, the boundary is uniformly populated with a density equal to half the surface coverage, and no motile particles are initialized in the bulk. 

In our model, the important parameters to vary are $Pe$ and $P_{\text{off}}$. The former sets the diffusivity of the active colloids while keeping chamber geometry and particle velocity constant, and the latter has a nice interpretation in terms of processivity (average run length before detachment) when considering a suspension of motor-bound particles. The other parameters $P_{\text{on}}$ and the Reynolds number $Re = \rho U L/\mu$ (with $\rho$ and $\mu$ the density and dynamic viscosity of water, respectively) are fixed throughout our simulations, with $Re$ in the viscous regime. Unless explicitly mentioned otherwise, we report all results in the paper with nondimensionalized units, $u^* = u/v_w$, $r^* = r/L$, $t^* = t/(L/v_w)$, and for simplicity we drop the asterisks.

\section*{Results}

\subsection*{Optimal transport and flow topology}

We first consider a simple benchmark flow structure: uniform motion to the right in a confined chamber, with one active and five no-slip surfaces (Fig.~\ref{Fig1}E-F, see also Movie S3).
The steady state on the boundary for one particular choice of parameters shows an accumulation of particles at the right wall (at the plus end of the actin filaments) and a depletion of particles on the left (Fig.~\ref{Fig1}E). The fluid at the right boundary is forced upwards, creating a steady state vortex in the $xz$-plane (Fig.~\ref{Fig1}F). Varying the P\'eclet number and detachment rates shows that higher streaming magnitudes occur for lower values of $Pe$ and an intermediate value of $P_\text{off}$ (Fig.~\ref{Fig1}G and Fig. S2; for a more detailed description of our choice in parameters and the parameter sweep range, refer to Sec. II of the SI). This optimum is explained as follows. On the one hand, attached particles will tend to accumulate at the chamber edges within a timescale $\tau_w \sim L/v_w$. Overly processive motors will therefore on average reach the opposite wall before they detach into the bulk, reducing the streaming velocity. This sets a lower bound on the detachment probability, $\tau_{detach} = 1/P_{\text{off}} \leq \tau_w$, so we require that $P_{\text{off}} \geq v_w/L$ in order to establish nontrivial streaming velocities in confined volumes. 
On the other hand, particles that are not processive enough do not spend enough time at the surface to contribute significant momentum injection. A large diffusion coefficient (i.e. small P\'eclet number) helps to offset particle accumulation at edges and also homogenises density fluctuations, which increases the streaming strength and stability, allowing for the establishment of steady-state flow structures (Movies S2-3).
This is maximised in the limit $Pe \to 0$, when diffusion dominates advection, as motors spread through the box uniformly. 

In summary, the flow velocity can be optimised with a high diffusivity and an intermediate processivity. Note that experimental realizations place constraints on these parameters, as discussed in Supplementary Information Section II. We further note that the properties of the phase space diagram are particular to the geometry we have considered, and other cases (such as periodic boundary conditions) may yield very different results (Fig. S2, Movie S1). 
Moreover, uniform motion in confined chambers will lead to recirculating streamlines that can transport particles in the $\hat{\vec{z}}$ direction, despite that activity on the surface is directed uniquely along $\hat{\vec{x}}$ (Fig.~\ref{Fig1}F).

\subsection*{Light-modulated surfaces}

\begin{figure*}
\includegraphics[width=0.95\textwidth]{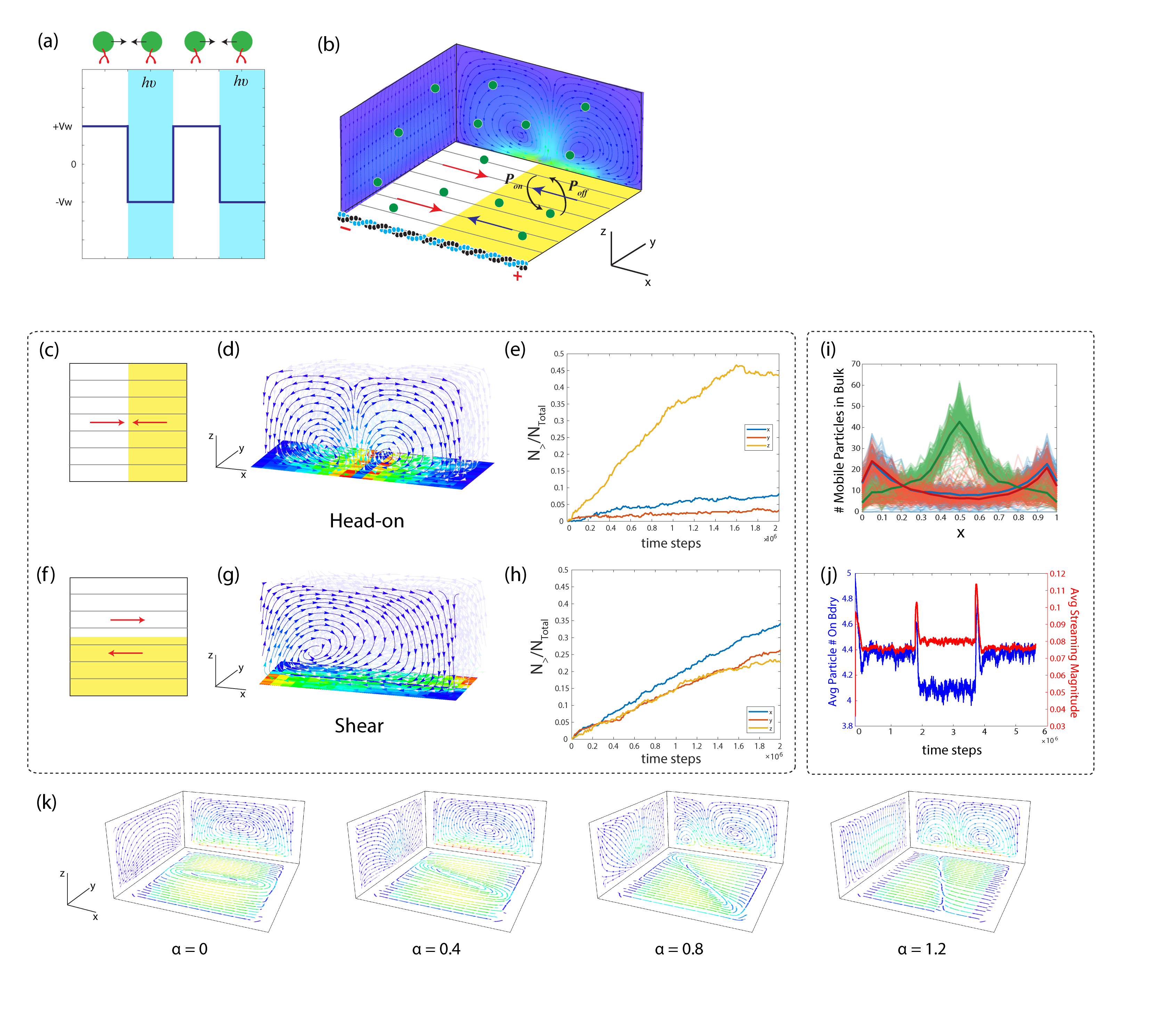}
\caption{\label{Fig2} 
Static and dynamic surface-patterned defects using light-controlled active colloidal particles in confined volumes. 
\textbf{(a)} A cartoon depiction of an active colloidal particle engineered to reverse direction along a track of fixed orientation upon illumination by light.  
\textbf{(b)} Schematic of a light-patterned boundary in a confined chamber of dimensions $N_x \times N_y \times N_z = 20 \times 20 \times 10$. The orientation of track filaments remains identical to Fig. \ref{Fig1}A, but now the $x>0.5$ half of the plane is illuminated by light. This leads to a line defect along $x=0.5$, where fluid moving in opposite directions collides and is pushed upward, creating two distinct vortices. 
\textbf{(c,f)} Light patterns giving rise to the head-on and shear defects, and 
\textbf{(d,g)} the emergent steady state flow structures. 
\textbf{(e,h)} The head-on and shear defects preferentially mix passively advecting particles ($Pe_{\text{tracer}}=10^{4}$) in different directions. Plotted on the $y$ axis is the fraction of tracer particles in the initially empty portion of the box as a function of time. Here $N_>$ denotes the number of particles at $x,y,z > 0.5, 0.5, 0.25$ for each of the three curves, respectively. 
\textbf{(e)} For the head-on defect, particles are nearly evenly distributed between the $z<0.25$ and $z> 0.25$ halves of the box, but mostly remain in the $x<0.5$ and $y<0.5$ regions, showing that there is little mixing in $x$ and $y$. 
\textbf{(h)} On the other hand, the shear defect is less effective at mixing in $\hat{z}$ but is able to mix particles along $\hat{y}$. 
\textbf{(i,j)} An optically controllable system allows easy temporal switching from one flow structure to another. As proof of principle, we switch between head-on and shear defects, with period $2 \times 10^6$ timesteps and parameters $P_{\text{off}} = 10^{-4}$ and $Pe=1$. 
\textbf{(i)} Shows the distribution of motile particles, where time moves from blue to green to red. 
\textbf{(j)} (In red) The average streaming magnitude, computed as the spatially averaged fluid magnitude as a function of time, and (in blue) the average number of particles attached to the boundary as a function of time. Interestingly, the average streaming magnitude of the head-on defect is greater even though the number of attached particles on the boundary is slightly less than that of the shear-defect. 
\textbf{(k)} Continuous perturbations to the flow structure: using light, we can transition continuously between a shear and a head-on defect by smoothly varying the parameter $\alpha$, which denotes the angle of the light pattern with the $y$-axis. 
}
\end{figure*}

One way to engineer different fluid structures would be to manipulate the orientation of tracks on a surface, an experimental perturbation made difficult by the fact that these tracks are often permanent when laid out. 
Advances in optogenetics have enabled the dynamic control of active materials through applying an external perturbation with light. For example, a new class of engineered molecular motors are capable of changing their direction of motion along a filament in response to a light signal, providing a mechanism to reprogram the same surface by changing how the motors interact with it \cite{nakamura2014remote}. We now explore this regime to see how we can program and pattern bulk flow with variable surface light patterns (Fig.~\ref{Fig2}A,B).]

Consider the same configuration as before, with all tracks oriented along the $\hat{\vec{x}}$ direction. Suppose the $x>0.5$ (right) half of the box is illuminated, redirecting the optically controllable particles in that region toward the minus end (Fig.~\ref{Fig2}C). We term the resulting structure the ``head-on'' defect, since now two populations of motile particles are literally walking into each other at a given line defect defined by the light pattern. This gives rise to two distinct vortices on either side of the domain junction (Fig.~\ref{Fig2}D). Similarly, we can also choose to illuminate the $y>0.5$ half of the box, giving rise to a ``shear'' defect configuration (Fig.~\ref{Fig2}F,G). On a surface patterned with uniformly oriented tracks, the head-on and shear defects are the two fundamental modes that one can pattern with light. 

These two flow structures have different properties that may be useful for different applications. Suppose, for example, that at subsequent stages of a chemical process two mixing procedures are needed. Figures~\ref{Fig2}E,H show that the shear and head-on defects preferentially mix tracer particles in different directions. The head-on defect is most effective at mixing along the $\hat{z}$-coordinate, as evidenced by the concentration profile of tracers initially in the bottom half of the box being spread uniformly along $\hat{z}$ by the end of the simulation (Fig.~\ref{Fig2}E). On the other hand, the shear defect is more effective at mixing along the $\hat{y}$ coordinate (Fig.~\ref{Fig2}H and Fig.~S3). Note that in both cases the direction of motion along the boundary is strictly in the $\hat{x}$ direction -- it is the geometry of the confined chamber (i.e. the location of walls and defects) that gives rise to recirculating streamlines. 

A significant advantage of light-controlled surface patterning is the ease of transitioning from one flow structure to another. As a proof of principle, we conducted a simulation (see Movie S4) where we transition from a shear defect to a head-on defect, and back to a shear defect again, with period $2 \times 10^6$ time steps, detachment rate $P_{\text{off}} = 10^{-4}$ per time step and $Pe = 1$ (the optimum in Fig.~\ref{Fig1}G, chosen for illustrating the phenomenon although experimental realizations would likely be at higher $Pe$; see Supplementary Table I). In our model, we assume that the behavior of the motile particles switch instantaneously with the external light pertubation. Figs.~\ref{Fig2}I,J show that with each transition, the flow relaxes to its unique steady state for each boundary condition after a short relaxation time. %set by $\text{max}\left( \frac{L}{v_w}, \frac{L^2}{D}, \frac{1}{P_\text{off}}, \frac{1}{P_\text{on}} \right)$). 
Interestingly, the shear defect generates a slightly lower streaming velocity despite having on average more particles attached to the boundary. Another characteristic of patterning with light is the ability to make continuous perturbations to the flow field. Fig.~\ref{Fig2}K depicts a continuous transformation from a shear to a head-on defect by smoothing varying the angle $\alpha$ of the light pattern with the $y$-axis, allowing for not only spatial but also precise temporal control of flow patterns (see also Movies S5-6). 

\subsection*{Systematic design of flow patterns in confined geometries}

Next, we consider the challenge of designing bulk flow patterns and consider the breadth of design space available in this problem.
To speed up our simulations, we continue in the optimal limit $Pe \to 0$ where the P\'eclet number tends to zero, when active particles cover all surfaces uniformly (or for a uniform carpet of cilia), and simulate the steady state flow structure in chambers of size $N_x \times N_y \times N_z = 40 \times 40 \times 40$. Hence, we no longer simulate the particle dynamics explicitly and study the steady-state flow structures obtained by patterning surfaces directly with constant slip velocities. This indeed limits the class of patterns that might be feasible, but allows us to understand the limit of vanishing P\'eclet number where active particles can reach surfaces in abundance. 
Though these surface velocities can be patterned arbitrarily, the resulting bulk flows must still obey the constraints set by the Stokes equations and incompressibility. The question arises how the optogenetic design can be optimized to alleviate those constraints in terms of transport and streamline connectivity.

\begin{figure*}
\includegraphics[width=\textwidth]{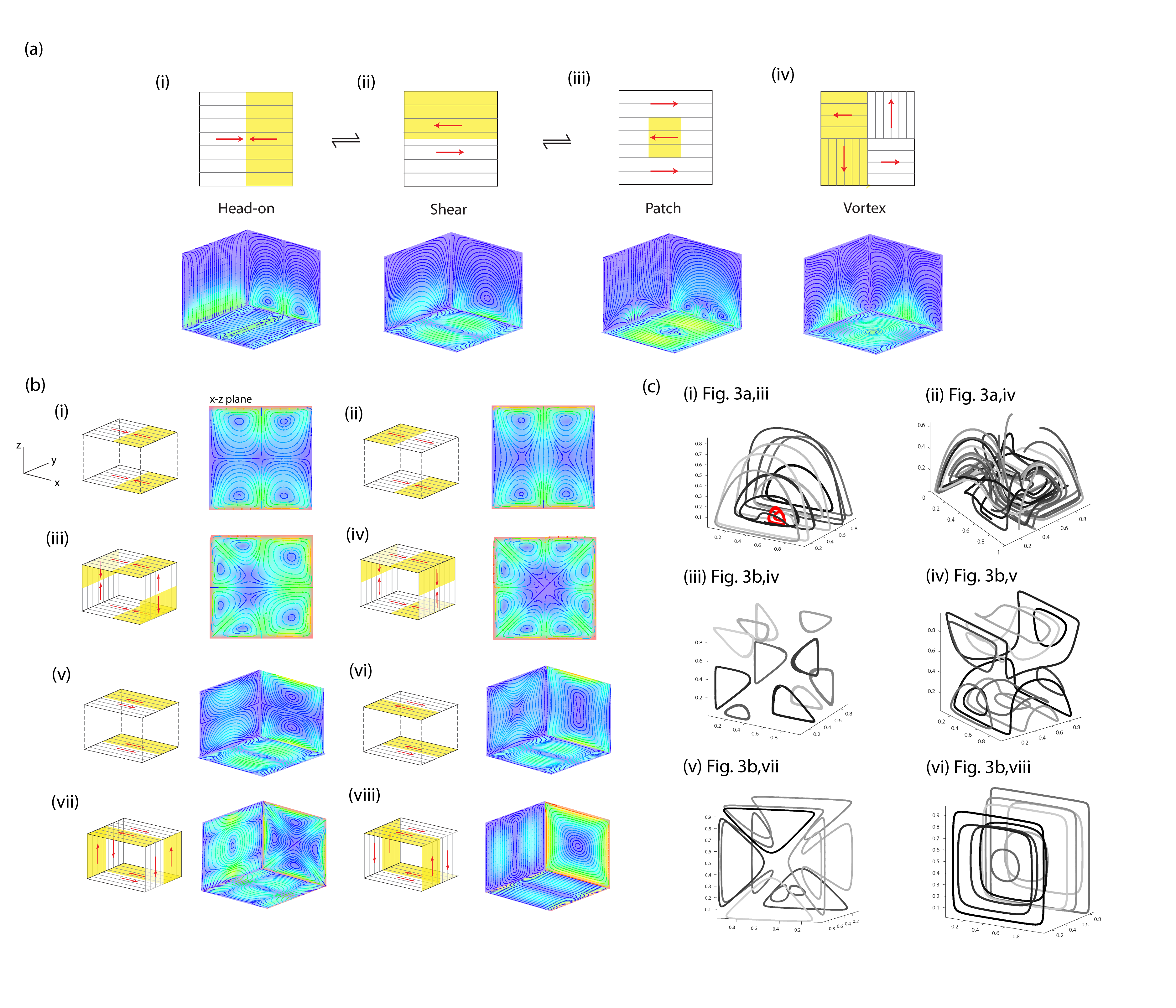}
\caption{\label{Fig3} 
Design space of surface-driven flows. All flow structures are solved on a $N_x \times N_y \times N_z = 40 \times 40 \times 40$ grid.
\textbf{(a)} Each panel depicts a pattern of light on a single active boundary (top) and the corresponding 3D flow structure (bottom). Note that panels i-iii are interchangeable by light, whereas panel iv introduces additional complexity where the orientation of actin on the surface is no longer fully uniform. 
\textbf{(b)} Panels i-iv consider patterning on either 2 or 4 surfaces with head-on defects. Since the resulting streamlines are largely 2D, only $xz$ cross sections of the flow structures are plotted. 
Panels vi-viii considers the same but with shear defects, and with the full 3D flow structure to highlight the recirculating streamlines. 
\textbf{(c)} Select streamlines from Fig. 3a-b are plotted to enhance visualization of the flow structures. Note the red-streamlines in panel i, which highlights the particles trapped in the clockwise vortex established by the oppositely-moving patch in Fig. 3a,iii. Contrasting panel iii with iv-v shows the effect of the recirculating streamlines -- particles in the former remain advected in the $xz$-plane whereas particles in the latter circulate in $xy$. Interestingly, however, panel vi returns to 2D streamlines despite that the boundaries are composed entirely of shear defects. 
}
\end{figure*}

For surface patterning, we utilize the language of defects that refer to zones where surface bound active particles dramatically change behavior. Starting with a single active surface with uniformly oriented filaments, different regions of left- or right-moving fluid can be created with light, as depicted for the head-on, shear, and patch defects (Fig.~\ref{Fig3}A, i-iii). Again, all three patterns are interchangeable by dynamically changing the pattern of light. We can add an additional handle by no longer subjecting the filaments to be uniformly oriented. Alternating orthogonal patches of tracks, in tandem with a light pattern on the surface, can give rise to flow structures such a vortex (Fig.~\ref{Fig3}A, iv). 

Integrated streamlines of the patch defect in Fig. ~\ref{Fig3}C,i highlights the separatrix formed by a small region of oppositely moving flow. Streamlines shown in red traverse clockwise (CW) and are centered on top of the patch, whereas all other streamlines travel counterclockwise (CCW). The head-on and patch defects therefore have a compartmentalizing effect, with regions of streamlines that do not mix. Conversely, Fig.~\ref{Fig3}C,iv shows that the streamlines of the vortex defect traverse the $xy$-plane as well as a distance of over half the height of the grid in $z$. The shear and vortex defects are therefore effective fluid mixers. These countervailing properties of compartmentalizing and mixing can guide the design of numerous functions useful in self-driven microfluidics context. 

We can further build upon the complexity of our designs by patterning multiple surfaces at once. Fig.~\ref{Fig3}B approaches this systematically by considering only head-on (i-iv) or shear defects (v-viii) on 2 or 4 surfaces. Interpreting the resultant flow structures created by head-on defects is straightforward: two stable vortices will form on either side of a defect, where the flow either moves toward ($\rightarrow \leftarrow$) or away ($\leftarrow \rightarrow$) from each other. Furthermore, the integrated streamlines of the 8 vortex structure (Fig.~\ref{Fig3}C,iii) shows that the streamlines are two dimensional. Each vortex can therefore be considered as its own compartment. 

On the other hand, patterning with shear defects can lead to mixing within each compartment (Fig. S4). Fig. ~\ref{Fig3}B,vii depicts four consecutive active surfaces, with each pair of opposite faces patterned with shear defects of opposite signs ($\uparrow \downarrow$ and $\downarrow \uparrow$). In the $xz$-plane, this gives rise to head-on defects at the four corners, creating four stable vortices. However, recirculation in the $yz$ and $xy$-planes cause the streamlines within the four vortices to traverse along the $y$-coordinate, as depicted in Fig.~\ref{Fig3}C,v. The expectation that shear defects give rise to three dimensional streamlines is a general but not very robust rule, however. Fig.~\ref{Fig3}B,viii depicts a surface pattern similar to Fig.~\ref{Fig3}B,vii, except with the patterns on the two faces with surface normal $\hat{x}$ flipped. The result is that the head-on defects in the $xz$-plane are completely eliminated, leading to a continuous current that runs CCW in the $y<0.5$ region of the box and CW in the other. Interestingly, the resulting streamlines (Fig.~\ref{Fig3}C,vi) are again co-planar.
%$\updownarrows$ and $\downuparrows)$
%*****

\subsection*{The interior Squirmer model}

Given the quickly increasing complexity of the resulting flows and the myriad possibilities of surface patterns, it is clear that adopting a more analytical approach to understanding the design space of boundary-driven flows is required. For this we turn to solutions of the squirmer model on a sphere \cite{blake1971spherical, reigh2017swimming, pedley2016spherical, pak2014generalized, fadda2020dynamics}. Initially adopted to study the external flow fields of microswimmers, we invert the problem and study instead the flow structure within the sphere \cite{happel2012low}, subject to the condition that flows normal to the surface vanish on the boundary.  
The resulting general solution for incompressible Stokes flow inside  sphere is: 

\begin{multline}
   u_r(r,\theta,\phi) = \sum_{n=1}^{\infty} \sum_{m=0}^n  nr^{n-1}P_n^m\left(1-\frac{r^2}{R^2}\right)\times \\ 
   (b_{mn} \cos m\phi + \tilde{b}_{mn}\sin m\phi),
\end{multline}

\begin{multline}
   u_\theta = \sum_{n=1}^{\infty} \sum_{m=0}^n r^{n+1} \sin\theta P_n^{m'} \left(\frac{n+3}{(n+1)R^2}-\frac{1}{r^2} \right)\times \\ (b_{mn} \cos m\phi + \tilde{b}_{mn} \sin m\phi) + \\ \frac{mr^n}{\sin\theta} P_n^m (\tilde{c}_{mn} \cos m \phi - c_{mn}\sin m \phi),
\end{multline}

\begin{multline}
   u_\phi = \sum_{n=1}^{\infty} \sum_{m=0}^n \frac{mr^{n+1} P_n^m}{\sin\theta} \left(\frac{1}{r^2} - \frac{(n+3)}{(n+1)R^2} \right) \times \\ (\tilde{b}_{mn}\cos m \phi - b_{mn} \sin m \phi) + \\ r^n\sin\theta P_n^{m'}(  c_{mn} \cos m \phi +  \tilde{c}_{mn}\sin m \phi),
\end{multline} 
where $P_m^n$ are the associated Legendre polynomials indexed by integers $m,n$, and $R$ the radius of the sphere. For the sake of simplicity we set $b_{mn} = \tilde{b}_{mn}$ and $c_{mn} = \tilde{c}_{mn}$, which fixes the phase of $\phi$ whilst keeping the topology the same. The modes are thus denoted by two variables, $b_{mn}$ and $c_{mn}$.

The first few axisymmetric modes are shown to match with the simulated flow structures on a grid (Fig.~\ref{Fig3}). We observe that the $b$-modes are aligned longitudinally across the sphere, whereas the $c$-modes run along lines of latitude and form closed streamlines. Cross sections of the box and sphere reveal the similarities between the internal flow structures. The topological equivalence between a sphere and a box dictates that these flow structures should be compatible. We do note, however, that corner effects can give rise to eddies that are unique to the geometry of a box \cite{LaugaStokesCorner}. Most interestingly, the $b_{20}$ and $c_{20}$ modes have built-in defects analogous to topologies proposed earlier: the former consists of a line of head-on defects at the equator, while the latter consists of two oppositely rotating hemispheres, giving rise to a shear defect along the equator. The spherical solutions therefore present a natural framework in which to embed the design space of surface-driven flow structures. 

Plots of higher order modes and their streamlines (Fig.~\ref{Fig5} and Figs.~S5-9) show that these general rules of thumb still apply: $b$ modes give rise to patches of oppositely moving flow on the surface, while $c$-modes give rise to closed vortices. Higher order modes give rise to more patches or more vortices, corresponding to smaller compartments in which tracer particles can traverse (Figs.~S8-9). 
The $b$-modes are better at mixing particles radially within their compartments due to streamlines that redirect particles toward the $z = 0$ axis, whereas particles move along concentric closed curves at approximately constant radius in $c$-modes (Fig. S5-7). These properties can be combined and used when designing microfluidic devices that require specific bulk flow patterns. 

\subsection*{Chaotic mixing by mode superposition}

Inspired by previous work showing that Stokes flows within droplets can give rise to chaotic streamlines \cite{bajer1990class,stone1991chaotic,ward2006chaotic}, we ask whether our surface-driven flow patterns can do the same. 
Indeed, we find evidence of chaotic mixing in the superposed $b$ and $c$ modes (Fig.~\ref{Fig5}). 
Whereas the individual $b$ and $c$ modes do not feature chaos, we do find evidence of chaotic mixing in superpositions of these modes (Fig.~\ref{Fig5}). 
To quantify this, we consider the trajectories of tracer particle pairs that are initially spaced a distance $dr_0 = 10^{-6}$ apart (in dimensionless units).
Fig.~\ref{Fig5}B depicts 10 such trajectories subject to the $b_{21}$ flow field only. 
The blue dots denote the starting positions of one pair of particles, which cannot be visually resolved.
The green dots denote their final positions after integrating to time $t=500$. The red curve in Fig.~\ref{Fig5}F plots the separation $dr$ as a function of time, averaged over 1000 randomly seeded trajectories, and shows that the particle pairs remain close together throughout their trajectories, with a final separation less than $dr=10^{-3}$. Similarly, the blue curve in Fig.~\ref{Fig5}F suggests that the $c_{21}$ mode is not chaotic. 

Trajectories of the $b_{21}+c_{21}$ modes combined, however, will on average diverge rapidly to a distance comparable to the system size $(R=1$). Fig.~\ref{Fig5}C depicts a single pair of trajectories that begin at the blue dot near the center of the sphere. After a time $t=500$, the green dots show the separation of these two particles at a distance comparable to the diameter of the sphere. 
This chaotic mixing is further illustrated by the Poincare sections at $x=0$ for the $b_{21}$ and $b_{21}+c_{21}$ modes (Fig.~\ref{Fig5}D-E), each built from 1000 randomly seeded trajectories. The Poincare section of $b_{21}$ is notably sparser than that of $b_{21}+c_{21}$, showing that the superposition of modes is far more effective at mixing.  

To understand this better, we map out the entire phase space of mixing potential by superpositions of the $n=2$ modes (Fig.~\ref{Fig5}G).
This heat map shows the exponent of $dr$ for each pair of superposed modes. Again, only the superposed modes show signs of mixing. 
We note, however, that only the modes consisting of $b$ superposed onto $c$ lead to chaotic advection, suggesting that fundamental properties from each mode are necessary ingredients. 
Somewhat surprisingly, even the axisymmetric mode $b_{20}$ showed moderate signs of mixing when superposed with the $c_{21}$ and $c_{22}$ modes, but the $b_{21}+c_{22}$ mode combination did not. 
Though we can predict the flow structures of simple surface patterns, it is not obvious, for example, which of the more complex topologies will lead to chaotic mixing and which do not. 
Prior theoretical work has shown that fluid properties such as stretching, twisting, and folding are essential for chaotic mixing \cite{bajer1990class,stone1991chaotic,ward2006chaotic,smith2019geometry,ottino1989kinematics}, but further work is required to understand how such concepts may arise in the devices proposed here.

\section*{Discussion}

In this paper, we have proposed a new class of flow patterns that utilize surface-driven flows rather than externally applied pressure gradients in confined geometries. In the first part of the paper,  we consider a realization of active boundaries using light-controllable surface-driven active particles and present the design space associated with bulk flows programmed by surface activity. We model a suspension of active colloidal particles, which can bind to the directional tracks grafted onto a closed chamber surfaces. 
We demonstrate that these currents can be optimised by tuning the particle attachment and detachment probabilities, and other physical parameters including the diffusivity and the surface velocity.
The particular geometry that we considered (uniform flow in confinement) required that diffusion be large relative to advection (low $Pe$ limit) in order to establish steady state flows, reducing the possible effects of flow-enhanced concentration gradients or density fluctuations. However, our present study has only explored the realm of low velocities and Reynolds number. Future work remains to be done to probe the effects of advective feedback at intermediate Re as well as other geometries.

The non-equilibrium transport by active particles is augmented further by optogenetic perturbations, using active particles that reverse velocity upon illumination.
Hence, different classes of topological defects are created by light patterning, giving rise to flow structures that each have different functions including hydrodynamic compartmentalisation, translation and rotation, and chaotic mixing.

Importantly, the concept of active boundaries is much more general than any specific implementation. 
In this paper we used the analogy of light-controlled active particles walking in one direction, but multiple channels can be combined to create a full orthogonal control of the active carpet. 
For example, myosin motors tuned by one wavelength of light can generate surface flows via actin tracks along the $x$ direction, while kinesins run on microtubules grafted along $y$ with control via another light bandwidth.
Moreover, besides molecular motors, these surface-driven flows may equally be driven by artificial cilia \cite{den2008artificial, van2009printed} in spatially patterned magnetic landscapes, engineered bacterial carpets \cite{darnton2004moving, mathijssen2019nutrient, jin2018biofilm}, hydrogel actuators, liquid crystal elastomers, and other responsive materials \cite{warner2007liquid, stuart2010emerging, wang2013light}. But regardless of implementation, the same fundamental modes and design principles can emerge. 

The generality of the concept of surface-driven flows led us to develop an analytic theory that reveals the possible flow structures in terms of fundamental modes, which may be superposed spatiotemporally. 
Hence, microfluidic flows may be designed at microscopic resolution without the need for physical channel fabrication.
The simplicity of such a microfluidic design platform is the minimal amount of experimental manipulation required during operation: a single active surface can already give rise to innumerable flow structures, which can be switched rapidly upon demand by direct spatial light modulation. Multiple active surfaces render this design space even richer. Hence, highly complex and dynamic time-varying protocols may be designed with these internally driven flows. Overall, this platform provides a fertile testing ground for understanding and designing active carpets from first principles.
 
\section*{Acknowledgments}
We acknowledge Paul Ruijgrok for many helpful discussion throughout this project and for detailed comments on the final manuscript. This work was supported by funding from the National Science Foundation Center for Cellular Construction (NSF grant DBI-1548297), a Human Frontier Science Program Fellowship to A.M. (LT001670/2017), and the Keck Foundation. 

\bibliography{biblio}

\begin{thebibliography}{56}%
\makeatletter
\providecommand \@ifxundefined [1]{%
 \@ifx{#1\undefined}
}%
\providecommand \@ifnum [1]{%
 \ifnum #1\expandafter \@firstoftwo
 \else \expandafter \@secondoftwo
 \fi
}%
\providecommand \@ifx [1]{%
 \ifx #1\expandafter \@firstoftwo
 \else \expandafter \@secondoftwo
 \fi
}%
\providecommand \natexlab [1]{#1}%
\providecommand \enquote  [1]{``#1''}%
\providecommand \bibnamefont  [1]{#1}%
\providecommand \bibfnamefont [1]{#1}%
\providecommand \citenamefont [1]{#1}%
\providecommand \href@noop [0]{\@secondoftwo}%
\providecommand \href [0]{\begingroup \@sanitize@url \@href}%
\providecommand \@href[1]{\@@startlink{#1}\@@href}%
\providecommand \@@href[1]{\endgroup#1\@@endlink}%
\providecommand \@sanitize@url [0]{\catcode `\\12\catcode `\$12\catcode
  `\&12\catcode `\#12\catcode `\^12\catcode `\_12\catcode `\%12\relax}%
\providecommand \@@startlink[1]{}%
\providecommand \@@endlink[0]{}%
\providecommand \url  [0]{\begingroup\@sanitize@url \@url }%
\providecommand \@url [1]{\endgroup\@href {#1}{\urlprefix }}%
\providecommand \urlprefix  [0]{URL }%
\providecommand \Eprint [0]{\href }%
\providecommand \doibase [0]{http://dx.doi.org/}%
\providecommand \selectlanguage [0]{\@gobble}%
\providecommand \bibinfo  [0]{\@secondoftwo}%
\providecommand \bibfield  [0]{\@secondoftwo}%
\providecommand \translation [1]{[#1]}%
\providecommand \BibitemOpen [0]{}%
\providecommand \bibitemStop [0]{}%
\providecommand \bibitemNoStop [0]{.\EOS\space}%
\providecommand \EOS [0]{\spacefactor3000\relax}%
\providecommand \BibitemShut  [1]{\csname bibitem#1\endcsname}%
\let\auto@bib@innerbib\@empty
%</preamble>
\bibitem [{\citenamefont {Ramaswamy}(2010)}]{sriram_review}%
  \BibitemOpen
  \bibfield  {author} {\bibinfo {author} {\bibfnamefont {S.}~\bibnamefont
  {Ramaswamy}},\ }\bibfield  {title} {\enquote {\bibinfo {title} {The mechanics
  and statistics of active matter},}\ }\href {\doibase
  10.1146/annurev-conmatphys-070909-104101} {\bibfield  {journal} {\bibinfo
  {journal} {Ann. Rev. Cond. Matt. Phys.}\ }\textbf {\bibinfo {volume} {1}},\
  \bibinfo {pages} {323--345} (\bibinfo {year} {2010})}\BibitemShut {NoStop}%
\bibitem [{\citenamefont {Marchetti}\ \emph {et~al.}(2013)\citenamefont
  {Marchetti}, \citenamefont {Joanny}, \citenamefont {Ramaswamy}, \citenamefont
  {Liverpool}, \citenamefont {Prost}, \citenamefont {Rao},\ and\ \citenamefont
  {Simha}}]{marchetti_review}%
  \BibitemOpen
  \bibfield  {author} {\bibinfo {author} {\bibfnamefont {M.~C.}\ \bibnamefont
  {Marchetti}}, \bibinfo {author} {\bibfnamefont {J.~F.}\ \bibnamefont
  {Joanny}}, \bibinfo {author} {\bibfnamefont {S.}~\bibnamefont {Ramaswamy}},
  \bibinfo {author} {\bibfnamefont {T.~B.}\ \bibnamefont {Liverpool}}, \bibinfo
  {author} {\bibfnamefont {J.}~\bibnamefont {Prost}}, \bibinfo {author}
  {\bibfnamefont {M.}~\bibnamefont {Rao}}, \ and\ \bibinfo {author}
  {\bibfnamefont {R.~A.}\ \bibnamefont {Simha}},\ }\bibfield  {title} {\enquote
  {\bibinfo {title} {Hydrodynamics of soft active matter},}\ }\href {\doibase
  10.1103/RevModPhys.85.1143} {\bibfield  {journal} {\bibinfo  {journal} {Rev.
  Mod. Phys.}\ }\textbf {\bibinfo {volume} {85}},\ \bibinfo {pages}
  {1143--1189} (\bibinfo {year} {2013})}\BibitemShut {NoStop}%
\bibitem [{\citenamefont {Doostmohammadi}\ \emph {et~al.}(2018)\citenamefont
  {Doostmohammadi}, \citenamefont {Ignés-Mullol}, \citenamefont {Yeomans},\
  and\ \citenamefont {Sagués}}]{activenematics_review}%
  \BibitemOpen
  \bibfield  {author} {\bibinfo {author} {\bibfnamefont {A.}~\bibnamefont
  {Doostmohammadi}}, \bibinfo {author} {\bibfnamefont {J.}~\bibnamefont
  {Ignés-Mullol}}, \bibinfo {author} {\bibfnamefont {J.}~\bibnamefont
  {Yeomans}}, \ and\ \bibinfo {author} {\bibfnamefont {F.}~\bibnamefont
  {Sagués}},\ }\bibfield  {title} {\enquote {\bibinfo {title} {Active
  nematics},}\ }\href {\doibase 10.1038/s41467-018-05666-8} {\bibfield
  {journal} {\bibinfo  {journal} {Nat. Commun.}\ }\textbf {\bibinfo {volume}
  {9}},\ \bibinfo {pages} {3246} (\bibinfo {year} {2018})}\BibitemShut
  {NoStop}%
\bibitem [{\citenamefont {Needleman}\ and\ \citenamefont
  {Dogic}(2017)}]{needleman_review}%
  \BibitemOpen
  \bibfield  {author} {\bibinfo {author} {\bibfnamefont {D.}~\bibnamefont
  {Needleman}}\ and\ \bibinfo {author} {\bibfnamefont {Z.}~\bibnamefont
  {Dogic}},\ }\bibfield  {title} {\enquote {\bibinfo {title} {Active matter at
  the interface between materials science and cell biology},}\ }\href {\doibase
  10.1038/natrevmats.2017.48} {\bibfield  {journal} {\bibinfo  {journal} {Nat.
  Rev. Mater.}\ }\textbf {\bibinfo {volume} {2}},\ \bibinfo {pages} {17048}
  (\bibinfo {year} {2017})}\BibitemShut {NoStop}%
\bibitem [{\citenamefont {Vicsek}\ \emph {et~al.}(1995)\citenamefont {Vicsek},
  \citenamefont {Czir\'ok}, \citenamefont {Ben-Jacob}, \citenamefont {Cohen},\
  and\ \citenamefont {Shochet}}]{vicsek1995}%
  \BibitemOpen
  \bibfield  {author} {\bibinfo {author} {\bibfnamefont {T.}~\bibnamefont
  {Vicsek}}, \bibinfo {author} {\bibfnamefont {A.}~\bibnamefont {Czir\'ok}},
  \bibinfo {author} {\bibfnamefont {E.}~\bibnamefont {Ben-Jacob}}, \bibinfo
  {author} {\bibfnamefont {I.}~\bibnamefont {Cohen}}, \ and\ \bibinfo {author}
  {\bibfnamefont {O.}~\bibnamefont {Shochet}},\ }\bibfield  {title} {\enquote
  {\bibinfo {title} {Novel type of phase transition in a system of self-driven
  particles},}\ }\href {\doibase 10.1103/PhysRevLett.75.1226} {\bibfield
  {journal} {\bibinfo  {journal} {Phys. Rev. Lett.}\ }\textbf {\bibinfo
  {volume} {75}},\ \bibinfo {pages} {1226--1229} (\bibinfo {year}
  {1995})}\BibitemShut {NoStop}%
\bibitem [{\citenamefont {Cavagna}\ \emph {et~al.}(2010)\citenamefont
  {Cavagna}, \citenamefont {Cimarelli}, \citenamefont {Giardina}, \citenamefont
  {Parisi}, \citenamefont {Santagati}, \citenamefont {Stefanini},\ and\
  \citenamefont {Viale}}]{flocks_scalefree}%
  \BibitemOpen
  \bibfield  {author} {\bibinfo {author} {\bibfnamefont {A.}~\bibnamefont
  {Cavagna}}, \bibinfo {author} {\bibfnamefont {A.}~\bibnamefont {Cimarelli}},
  \bibinfo {author} {\bibfnamefont {I.}~\bibnamefont {Giardina}}, \bibinfo
  {author} {\bibfnamefont {G.}~\bibnamefont {Parisi}}, \bibinfo {author}
  {\bibfnamefont {R.}~\bibnamefont {Santagati}}, \bibinfo {author}
  {\bibfnamefont {F.}~\bibnamefont {Stefanini}}, \ and\ \bibinfo {author}
  {\bibfnamefont {M.}~\bibnamefont {Viale}},\ }\bibfield  {title} {\enquote
  {\bibinfo {title} {Scale-free correlations in starling flocks},}\ }\href
  {\doibase 10.1073/pnas.1005766107} {\bibfield  {journal} {\bibinfo  {journal}
  {Proc. Nat. Acad. Sci.}\ }\textbf {\bibinfo {volume} {107}},\ \bibinfo
  {pages} {11865--11870} (\bibinfo {year} {2010})}\BibitemShut {NoStop}%
\bibitem [{\citenamefont {Toner}\ \emph {et~al.}(2005)\citenamefont {Toner},
  \citenamefont {Tu},\ and\ \citenamefont
  {Ramaswamy}}]{toner2005hydrodynamics}%
  \BibitemOpen
  \bibfield  {author} {\bibinfo {author} {\bibfnamefont {J.}~\bibnamefont
  {Toner}}, \bibinfo {author} {\bibfnamefont {Y.}~\bibnamefont {Tu}}, \ and\
  \bibinfo {author} {\bibfnamefont {S.}~\bibnamefont {Ramaswamy}},\ }\bibfield
  {title} {\enquote {\bibinfo {title} {Hydrodynamics and phases of flocks},}\
  }\href {\doibase 10.1016/j.aop.2005.04.011} {\bibfield  {journal} {\bibinfo
  {journal} {Ann. Phys.}\ }\textbf {\bibinfo {volume} {318}},\ \bibinfo {pages}
  {170--244} (\bibinfo {year} {2005})}\BibitemShut {NoStop}%
\bibitem [{\citenamefont {Sliusarenko}\ \emph {et~al.}(2007)\citenamefont
  {Sliusarenko}, \citenamefont {Zusman},\ and\ \citenamefont
  {Oster}}]{sliusarenko2007aggregation}%
  \BibitemOpen
  \bibfield  {author} {\bibinfo {author} {\bibfnamefont {O.}~\bibnamefont
  {Sliusarenko}}, \bibinfo {author} {\bibfnamefont {D.~R.}\ \bibnamefont
  {Zusman}}, \ and\ \bibinfo {author} {\bibfnamefont {G.}~\bibnamefont
  {Oster}},\ }\bibfield  {title} {\enquote {\bibinfo {title} {Aggregation
  during fruiting body formation in myxococcus xanthus is driven by reducing
  cell movement},}\ }\href {\doibase 10.1128/JB.01206-06} {\bibfield  {journal}
  {\bibinfo  {journal} {J. Bacteriol.}\ }\textbf {\bibinfo {volume} {189}},\
  \bibinfo {pages} {611--619} (\bibinfo {year} {2007})}\BibitemShut {NoStop}%
\bibitem [{\citenamefont {Wu}\ \emph {et~al.}(2009)\citenamefont {Wu},
  \citenamefont {Kaiser}, \citenamefont {Jiang},\ and\ \citenamefont
  {Alber}}]{wu2009periodic}%
  \BibitemOpen
  \bibfield  {author} {\bibinfo {author} {\bibfnamefont {Y.}~\bibnamefont
  {Wu}}, \bibinfo {author} {\bibfnamefont {A.~D.}\ \bibnamefont {Kaiser}},
  \bibinfo {author} {\bibfnamefont {Y.}~\bibnamefont {Jiang}}, \ and\ \bibinfo
  {author} {\bibfnamefont {M.~S.}\ \bibnamefont {Alber}},\ }\bibfield  {title}
  {\enquote {\bibinfo {title} {Periodic reversal of direction allows
  myxobacteria to swarm},}\ }\href {\doibase 10.1073/pnas.0811662106}
  {\bibfield  {journal} {\bibinfo  {journal} {Proc. Nat. Acad. Sci.}\ }\textbf
  {\bibinfo {volume} {106}},\ \bibinfo {pages} {1222--1227} (\bibinfo {year}
  {2009})}\BibitemShut {NoStop}%
\bibitem [{\citenamefont {Thutupalli}\ \emph {et~al.}(2015)\citenamefont
  {Thutupalli}, \citenamefont {Sun}, \citenamefont {Bunyak}, \citenamefont
  {Palaniappan},\ and\ \citenamefont {Shaevitz}}]{shaevitz2015}%
  \BibitemOpen
  \bibfield  {author} {\bibinfo {author} {\bibfnamefont {S.}~\bibnamefont
  {Thutupalli}}, \bibinfo {author} {\bibfnamefont {M.}~\bibnamefont {Sun}},
  \bibinfo {author} {\bibfnamefont {F.}~\bibnamefont {Bunyak}}, \bibinfo
  {author} {\bibfnamefont {K.}~\bibnamefont {Palaniappan}}, \ and\ \bibinfo
  {author} {\bibfnamefont {J.~W.}\ \bibnamefont {Shaevitz}},\ }\bibfield
  {title} {\enquote {\bibinfo {title} {Directional reversals enable myxococcus
  xanthus cells to produce collective one-dimensional streams during
  fruiting-body formation},}\ }\href {\doibase 10.1098/rsif.2015.0049}
  {\bibfield  {journal} {\bibinfo  {journal} {J. Roy. Soc. Interface}\ }\textbf
  {\bibinfo {volume} {12}},\ \bibinfo {pages} {20150049} (\bibinfo {year}
  {2015})}\BibitemShut {NoStop}%
\bibitem [{\citenamefont {Wensink}\ \emph {et~al.}(2012)\citenamefont
  {Wensink}, \citenamefont {Dunkel}, \citenamefont {Heidenreich}, \citenamefont
  {Drescher}, \citenamefont {Goldstein}, \citenamefont {L{\"o}wen},\ and\
  \citenamefont {Yeomans}}]{2012bsub_turbulence}%
  \BibitemOpen
  \bibfield  {author} {\bibinfo {author} {\bibfnamefont {H.~H.}\ \bibnamefont
  {Wensink}}, \bibinfo {author} {\bibfnamefont {J.}~\bibnamefont {Dunkel}},
  \bibinfo {author} {\bibfnamefont {S.}~\bibnamefont {Heidenreich}}, \bibinfo
  {author} {\bibfnamefont {K.}~\bibnamefont {Drescher}}, \bibinfo {author}
  {\bibfnamefont {R.~E.}\ \bibnamefont {Goldstein}}, \bibinfo {author}
  {\bibfnamefont {H.}~\bibnamefont {L{\"o}wen}}, \ and\ \bibinfo {author}
  {\bibfnamefont {J.~M.}\ \bibnamefont {Yeomans}},\ }\bibfield  {title}
  {\enquote {\bibinfo {title} {Meso-scale turbulence in living fluids},}\
  }\href {\doibase 10.1073/pnas.1202032109} {\bibfield  {journal} {\bibinfo
  {journal} {Proc. Nat. Acad. Sci.}\ }\textbf {\bibinfo {volume} {109}},\
  \bibinfo {pages} {14308--14313} (\bibinfo {year} {2012})}\BibitemShut
  {NoStop}%
\bibitem [{\citenamefont {Mathijssen}\ \emph {et~al.}(2019)\citenamefont
  {Mathijssen}, \citenamefont {Culver}, \citenamefont {Bhamla},\ and\
  \citenamefont {Prakash}}]{mathijssen2019collective}%
  \BibitemOpen
  \bibfield  {author} {\bibinfo {author} {\bibfnamefont {A.~J. T.~M.}\
  \bibnamefont {Mathijssen}}, \bibinfo {author} {\bibfnamefont
  {J.}~\bibnamefont {Culver}}, \bibinfo {author} {\bibfnamefont {M.~S.}\
  \bibnamefont {Bhamla}}, \ and\ \bibinfo {author} {\bibfnamefont
  {M.}~\bibnamefont {Prakash}},\ }\bibfield  {title} {\enquote {\bibinfo
  {title} {Collective intercellular communication through ultra-fast
  hydrodynamic trigger waves},}\ }\href {\doibase 10.1038/s41586-019-1387-9}
  {\bibfield  {journal} {\bibinfo  {journal} {Nature}\ }\textbf {\bibinfo
  {volume} {571}},\ \bibinfo {pages} {560--565} (\bibinfo {year}
  {2019})}\BibitemShut {NoStop}%
\bibitem [{\citenamefont {Woodhouse}\ and\ \citenamefont
  {Dunkel}(2017)}]{woodhouse2017active}%
  \BibitemOpen
  \bibfield  {author} {\bibinfo {author} {\bibfnamefont {F.~G.}\ \bibnamefont
  {Woodhouse}}\ and\ \bibinfo {author} {\bibfnamefont {J.}~\bibnamefont
  {Dunkel}},\ }\bibfield  {title} {\enquote {\bibinfo {title} {Active matter
  logic for autonomous microfluidics},}\ }\href {\doibase
  doi:10.1038/ncomms15169} {\bibfield  {journal} {\bibinfo  {journal} {Nat.
  Commun.}\ }\textbf {\bibinfo {volume} {8}},\ \bibinfo {pages} {15169}
  (\bibinfo {year} {2017})}\BibitemShut {NoStop}%
\bibitem [{\citenamefont {Woodhouse}\ and\ \citenamefont
  {Goldstein}(2012)}]{woodhouse2012spontaneous}%
  \BibitemOpen
  \bibfield  {author} {\bibinfo {author} {\bibfnamefont {F.~G.}\ \bibnamefont
  {Woodhouse}}\ and\ \bibinfo {author} {\bibfnamefont {R.~E.}\ \bibnamefont
  {Goldstein}},\ }\bibfield  {title} {\enquote {\bibinfo {title} {Spontaneous
  circulation of confined active suspensions},}\ }\href {\doibase
  10.1103/PhysRevLett.109.168105} {\bibfield  {journal} {\bibinfo  {journal}
  {Phys. Rev. Lett.}\ }\textbf {\bibinfo {volume} {109}},\ \bibinfo {pages}
  {168105} (\bibinfo {year} {2012})}\BibitemShut {NoStop}%
\bibitem [{\citenamefont {Goldstein}\ \emph {et~al.}(2008)\citenamefont
  {Goldstein}, \citenamefont {Tuval},\ and\ \citenamefont {van~de
  Meent}}]{goldstein2008microfluidics}%
  \BibitemOpen
  \bibfield  {author} {\bibinfo {author} {\bibfnamefont {R.~E.}\ \bibnamefont
  {Goldstein}}, \bibinfo {author} {\bibfnamefont {I.}~\bibnamefont {Tuval}}, \
  and\ \bibinfo {author} {\bibfnamefont {J.-W.}\ \bibnamefont {van~de Meent}},\
  }\bibfield  {title} {\enquote {\bibinfo {title} {Microfluidics of cytoplasmic
  streaming and its implications for intracellular transport},}\ }\href
  {\doibase 10.1073/pnas.0707223105} {\bibfield  {journal} {\bibinfo  {journal}
  {Proc. Nat. Acad. Sci.}\ }\textbf {\bibinfo {volume} {105}},\ \bibinfo
  {pages} {3663--3667} (\bibinfo {year} {2008})}\BibitemShut {NoStop}%
\bibitem [{\citenamefont {Ross}\ \emph {et~al.}(2019)\citenamefont {Ross},
  \citenamefont {Lee}, \citenamefont {Qu}, \citenamefont {Banks}, \citenamefont
  {Phillips},\ and\ \citenamefont {Thomson}}]{ross2018}%
  \BibitemOpen
  \bibfield  {author} {\bibinfo {author} {\bibfnamefont {T.~D.}\ \bibnamefont
  {Ross}}, \bibinfo {author} {\bibfnamefont {H.~J.}\ \bibnamefont {Lee}},
  \bibinfo {author} {\bibfnamefont {Z.}~\bibnamefont {Qu}}, \bibinfo {author}
  {\bibfnamefont {R.~A.}\ \bibnamefont {Banks}}, \bibinfo {author}
  {\bibfnamefont {R.}~\bibnamefont {Phillips}}, \ and\ \bibinfo {author}
  {\bibfnamefont {M.}~\bibnamefont {Thomson}},\ }\bibfield  {title} {\enquote
  {\bibinfo {title} {Controlling organization and forces in active matter
  through optically defined boundaries},}\ }\href {\doibase
  10.1038/s41586-019-1447-1} {\bibfield  {journal} {\bibinfo  {journal}
  {Nature}\ }\textbf {\bibinfo {volume} {572}},\ \bibinfo {pages} {224--229}
  (\bibinfo {year} {2019})}\BibitemShut {NoStop}%
\bibitem [{\citenamefont {Bustamante-Marin}\ and\ \citenamefont
  {Ostrowski}(2017)}]{ciliaclearance}%
  \BibitemOpen
  \bibfield  {author} {\bibinfo {author} {\bibfnamefont {X.~M.}\ \bibnamefont
  {Bustamante-Marin}}\ and\ \bibinfo {author} {\bibfnamefont {L.~E.}\
  \bibnamefont {Ostrowski}},\ }\bibfield  {title} {\enquote {\bibinfo {title}
  {Cilia and mucociliary clearance},}\ }\href {\doibase
  10.1101/cshperspect.a028241} {\bibfield  {journal} {\bibinfo  {journal}
  {{CSH} Persp. Biol.}\ }\textbf {\bibinfo {volume} {9}},\ \bibinfo {pages}
  {a028241} (\bibinfo {year} {2017})}\BibitemShut {NoStop}%
\bibitem [{\citenamefont {Elgeti}\ and\ \citenamefont
  {Gompper}(2013)}]{elgeti2013metachronal}%
  \BibitemOpen
  \bibfield  {author} {\bibinfo {author} {\bibfnamefont {J.}~\bibnamefont
  {Elgeti}}\ and\ \bibinfo {author} {\bibfnamefont {G.}~\bibnamefont
  {Gompper}},\ }\bibfield  {title} {\enquote {\bibinfo {title} {Emergence of
  metachronal waves in cilia arrays},}\ }\href {\doibase
  10.1073/pnas.1218869110} {\bibfield  {journal} {\bibinfo  {journal} {Proc.
  Nat. Acad. Sci.}\ }\textbf {\bibinfo {volume} {110}},\ \bibinfo {pages}
  {4470--4475} (\bibinfo {year} {2013})}\BibitemShut {NoStop}%
\bibitem [{\citenamefont {Goldestein}\ and\ \citenamefont
  {Meent}(2015)}]{goldstein_review}%
  \BibitemOpen
  \bibfield  {author} {\bibinfo {author} {\bibfnamefont {R.}~\bibnamefont
  {Goldestein}}\ and\ \bibinfo {author} {\bibfnamefont {J.}~\bibnamefont
  {Meent}},\ }\bibfield  {title} {\enquote {\bibinfo {title} {A physical
  perspective on cytoplasmic streaming},}\ }\href {\doibase
  10.1098/rsfs.2015.0030} {\bibfield  {journal} {\bibinfo  {journal} {Interface
  Focus}\ }\textbf {\bibinfo {volume} {5}},\ \bibinfo {pages} {20150030}
  (\bibinfo {year} {2015})}\BibitemShut {NoStop}%
\bibitem [{\citenamefont {Woodhouse}\ and\ \citenamefont
  {Goldstein}(2013)}]{woodhouse2013cytoplasmic}%
  \BibitemOpen
  \bibfield  {author} {\bibinfo {author} {\bibfnamefont {F.~G.}\ \bibnamefont
  {Woodhouse}}\ and\ \bibinfo {author} {\bibfnamefont {R.~E.}\ \bibnamefont
  {Goldstein}},\ }\bibfield  {title} {\enquote {\bibinfo {title} {Cytoplasmic
  streaming in plant cells emerges naturally by microfilament
  self-organization},}\ }\href {\doibase 10.1073/pnas.1302736110} {\bibfield
  {journal} {\bibinfo  {journal} {Proc. Nat. Acad. Sci.}\ }\textbf {\bibinfo
  {volume} {110}},\ \bibinfo {pages} {14132--14137} (\bibinfo {year}
  {2013})}\BibitemShut {NoStop}%
\bibitem [{\citenamefont {den Toonder}\ \emph {et~al.}(2008)\citenamefont {den
  Toonder}, \citenamefont {Bos}, \citenamefont {Broer}, \citenamefont
  {Filippini}, \citenamefont {Gillies}, \citenamefont {de~Goede}, \citenamefont
  {Mol}, \citenamefont {Reijme}, \citenamefont {Talen}, \citenamefont
  {Wilderbeek} \emph {et~al.}}]{den2008artificial}%
  \BibitemOpen
  \bibfield  {author} {\bibinfo {author} {\bibfnamefont {J.}~\bibnamefont {den
  Toonder}}, \bibinfo {author} {\bibfnamefont {F.}~\bibnamefont {Bos}},
  \bibinfo {author} {\bibfnamefont {D.}~\bibnamefont {Broer}}, \bibinfo
  {author} {\bibfnamefont {L.}~\bibnamefont {Filippini}}, \bibinfo {author}
  {\bibfnamefont {M.}~\bibnamefont {Gillies}}, \bibinfo {author} {\bibfnamefont
  {J.}~\bibnamefont {de~Goede}}, \bibinfo {author} {\bibfnamefont
  {T.}~\bibnamefont {Mol}}, \bibinfo {author} {\bibfnamefont {M.}~\bibnamefont
  {Reijme}}, \bibinfo {author} {\bibfnamefont {W.}~\bibnamefont {Talen}},
  \bibinfo {author} {\bibfnamefont {H.}~\bibnamefont {Wilderbeek}},  \emph
  {et~al.},\ }\bibfield  {title} {\enquote {\bibinfo {title} {Artificial cilia
  for active micro-fluidic mixing},}\ }\href {\doibase 10.1039/B717681C}
  {\bibfield  {journal} {\bibinfo  {journal} {Lab Chip}\ }\textbf {\bibinfo
  {volume} {8}},\ \bibinfo {pages} {533--541} (\bibinfo {year}
  {2008})}\BibitemShut {NoStop}%
\bibitem [{\citenamefont {Van~Oosten}\ \emph {et~al.}(2009)\citenamefont
  {Van~Oosten}, \citenamefont {Bastiaansen},\ and\ \citenamefont
  {Broer}}]{van2009printed}%
  \BibitemOpen
  \bibfield  {author} {\bibinfo {author} {\bibfnamefont {C.~L.}\ \bibnamefont
  {Van~Oosten}}, \bibinfo {author} {\bibfnamefont {C.~W.~M.}\ \bibnamefont
  {Bastiaansen}}, \ and\ \bibinfo {author} {\bibfnamefont {D.~J.}\ \bibnamefont
  {Broer}},\ }\bibfield  {title} {\enquote {\bibinfo {title} {Printed
  artificial cilia from liquid-crystal network actuators modularly driven by
  light},}\ }\href {\doibase 10.1038/nmat2487} {\bibfield  {journal} {\bibinfo
  {journal} {Nat. Mater.}\ }\textbf {\bibinfo {volume} {8}},\ \bibinfo {pages}
  {677} (\bibinfo {year} {2009})}\BibitemShut {NoStop}%
\bibitem [{\citenamefont {Bricard}\ \emph
  {et~al.}(2015{\natexlab{a}})\citenamefont {Bricard}, \citenamefont {Caussin},
  \citenamefont {Das}, \citenamefont {Savoie}, \citenamefont {Chikkadi},
  \citenamefont {Shitara}, \citenamefont {Chepizhko}, \citenamefont {Peruani},
  \citenamefont {Saintillan},\ and\ \citenamefont
  {Bartolo}}]{bricard2015emergent}%
  \BibitemOpen
  \bibfield  {author} {\bibinfo {author} {\bibfnamefont {A.}~\bibnamefont
  {Bricard}}, \bibinfo {author} {\bibfnamefont {J.-B.}\ \bibnamefont
  {Caussin}}, \bibinfo {author} {\bibfnamefont {D.}~\bibnamefont {Das}},
  \bibinfo {author} {\bibfnamefont {C.}~\bibnamefont {Savoie}}, \bibinfo
  {author} {\bibfnamefont {V.}~\bibnamefont {Chikkadi}}, \bibinfo {author}
  {\bibfnamefont {K.}~\bibnamefont {Shitara}}, \bibinfo {author} {\bibfnamefont
  {O.}~\bibnamefont {Chepizhko}}, \bibinfo {author} {\bibfnamefont
  {F.}~\bibnamefont {Peruani}}, \bibinfo {author} {\bibfnamefont
  {D.}~\bibnamefont {Saintillan}}, \ and\ \bibinfo {author} {\bibfnamefont
  {D.}~\bibnamefont {Bartolo}},\ }\bibfield  {title} {\enquote {\bibinfo
  {title} {Emergent vortices in populations of colloidal rollers},}\ }\href
  {\doibase 10.1038/ncomms8470} {\bibfield  {journal} {\bibinfo  {journal}
  {Nat. Comm.}\ }\textbf {\bibinfo {volume} {6}},\ \bibinfo {pages} {7470}
  (\bibinfo {year} {2015}{\natexlab{a}})}\BibitemShut {NoStop}%
\bibitem [{\citenamefont {Bechinger}\ \emph {et~al.}(2016)\citenamefont
  {Bechinger}, \citenamefont {Di~Leonardo}, \citenamefont {L{\"o}wen},
  \citenamefont {Reichhardt}, \citenamefont {Volpe},\ and\ \citenamefont
  {Volpe}}]{bechinger2016active}%
  \BibitemOpen
  \bibfield  {author} {\bibinfo {author} {\bibfnamefont {C.}~\bibnamefont
  {Bechinger}}, \bibinfo {author} {\bibfnamefont {R.}~\bibnamefont
  {Di~Leonardo}}, \bibinfo {author} {\bibfnamefont {H.}~\bibnamefont
  {L{\"o}wen}}, \bibinfo {author} {\bibfnamefont {C.}~\bibnamefont
  {Reichhardt}}, \bibinfo {author} {\bibfnamefont {G.}~\bibnamefont {Volpe}}, \
  and\ \bibinfo {author} {\bibfnamefont {G.}~\bibnamefont {Volpe}},\ }\bibfield
   {title} {\enquote {\bibinfo {title} {Active particles in complex and crowded
  environments},}\ }\href {\doibase 10.1103/RevModPhys.88.045006} {\bibfield
  {journal} {\bibinfo  {journal} {Rev. Mod. Phys.}\ }\textbf {\bibinfo {volume}
  {88}},\ \bibinfo {pages} {045006} (\bibinfo {year} {2016})}\BibitemShut
  {NoStop}%
\bibitem [{\citenamefont {Z{\"o}ttl}\ and\ \citenamefont
  {Stark}(2016)}]{zottl2016emergent}%
  \BibitemOpen
  \bibfield  {author} {\bibinfo {author} {\bibfnamefont {A.}~\bibnamefont
  {Z{\"o}ttl}}\ and\ \bibinfo {author} {\bibfnamefont {H.}~\bibnamefont
  {Stark}},\ }\bibfield  {title} {\enquote {\bibinfo {title} {Emergent behavior
  in active colloids},}\ }\href {\doibase 10.1088/0953-8984/28/25/253001}
  {\bibfield  {journal} {\bibinfo  {journal} {J. Phys. Cond. Mat.}\ }\textbf
  {\bibinfo {volume} {28}},\ \bibinfo {pages} {253001} (\bibinfo {year}
  {2016})}\BibitemShut {NoStop}%
\bibitem [{\citenamefont {Maass}\ \emph {et~al.}(2016)\citenamefont {Maass},
  \citenamefont {Kr{\"u}ger}, \citenamefont {Herminghaus},\ and\ \citenamefont
  {Bahr}}]{maass2016swimming}%
  \BibitemOpen
  \bibfield  {author} {\bibinfo {author} {\bibfnamefont {C.~C.}\ \bibnamefont
  {Maass}}, \bibinfo {author} {\bibfnamefont {C.}~\bibnamefont {Kr{\"u}ger}},
  \bibinfo {author} {\bibfnamefont {S.}~\bibnamefont {Herminghaus}}, \ and\
  \bibinfo {author} {\bibfnamefont {C.}~\bibnamefont {Bahr}},\ }\bibfield
  {title} {\enquote {\bibinfo {title} {Swimming droplets},}\ }\href {\doibase
  10.1146/annurev-conmatphys-031115-011517} {\bibfield  {journal} {\bibinfo
  {journal} {Ann. Rev. Cond. Mat. Phys.}\ }\textbf {\bibinfo {volume} {7}},\
  \bibinfo {pages} {171--193} (\bibinfo {year} {2016})}\BibitemShut {NoStop}%
\bibitem [{\citenamefont {Bricard}\ \emph
  {et~al.}(2015{\natexlab{b}})\citenamefont {Bricard}, \citenamefont {Caussin},
  \citenamefont {Das}, \citenamefont {Savoie}, \citenamefont {Chikkadi},
  \citenamefont {Shitara}, \citenamefont {Chepizhko}, \citenamefont {Peruani},
  \citenamefont {Saintillan},\ and\ \citenamefont {Bartolo}}]{quincke}%
  \BibitemOpen
  \bibfield  {author} {\bibinfo {author} {\bibfnamefont {A.}~\bibnamefont
  {Bricard}}, \bibinfo {author} {\bibfnamefont {J.-B.}\ \bibnamefont
  {Caussin}}, \bibinfo {author} {\bibfnamefont {D.}~\bibnamefont {Das}},
  \bibinfo {author} {\bibfnamefont {C.}~\bibnamefont {Savoie}}, \bibinfo
  {author} {\bibfnamefont {V.}~\bibnamefont {Chikkadi}}, \bibinfo {author}
  {\bibfnamefont {K.}~\bibnamefont {Shitara}}, \bibinfo {author} {\bibfnamefont
  {O.}~\bibnamefont {Chepizhko}}, \bibinfo {author} {\bibfnamefont
  {F.}~\bibnamefont {Peruani}}, \bibinfo {author} {\bibfnamefont
  {D.}~\bibnamefont {Saintillan}}, \ and\ \bibinfo {author} {\bibfnamefont
  {D.}~\bibnamefont {Bartolo}},\ }\bibfield  {title} {\enquote {\bibinfo
  {title} {Emergent vortices in populations of colloidal rollers},}\ }\href
  {\doibase 10.1038/ncomms8470} {\bibfield  {journal} {\bibinfo  {journal}
  {Nat. Commun.}\ }\textbf {\bibinfo {volume} {6}},\ \bibinfo {pages} {7470}
  (\bibinfo {year} {2015}{\natexlab{b}})}\BibitemShut {NoStop}%
\bibitem [{\citenamefont {Darnton}\ \emph {et~al.}(2004)\citenamefont
  {Darnton}, \citenamefont {Turner}, \citenamefont {Breuer},\ and\
  \citenamefont {Berg}}]{darnton2004moving}%
  \BibitemOpen
  \bibfield  {author} {\bibinfo {author} {\bibfnamefont {N.}~\bibnamefont
  {Darnton}}, \bibinfo {author} {\bibfnamefont {L.}~\bibnamefont {Turner}},
  \bibinfo {author} {\bibfnamefont {K.}~\bibnamefont {Breuer}}, \ and\ \bibinfo
  {author} {\bibfnamefont {H.~C.}\ \bibnamefont {Berg}},\ }\bibfield  {title}
  {\enquote {\bibinfo {title} {Moving fluid with bacterial carpets},}\ }\href
  {\doibase 10.1016/S0006-3495(04)74253-8} {\bibfield  {journal} {\bibinfo
  {journal} {Biophys. J.}\ }\textbf {\bibinfo {volume} {86}},\ \bibinfo {pages}
  {1863--1870} (\bibinfo {year} {2004})}\BibitemShut {NoStop}%
\bibitem [{\citenamefont {Mathijssen}\ \emph {et~al.}(2018)\citenamefont
  {Mathijssen}, \citenamefont {Guzm\'an-Lastra}, \citenamefont {Kaiser},\ and\
  \citenamefont {L\"owen}}]{mathijssen2019nutrient}%
  \BibitemOpen
  \bibfield  {author} {\bibinfo {author} {\bibfnamefont {A.~J. T.~M.}\
  \bibnamefont {Mathijssen}}, \bibinfo {author} {\bibfnamefont
  {F.}~\bibnamefont {Guzm\'an-Lastra}}, \bibinfo {author} {\bibfnamefont
  {A.}~\bibnamefont {Kaiser}}, \ and\ \bibinfo {author} {\bibfnamefont
  {H.}~\bibnamefont {L\"owen}},\ }\bibfield  {title} {\enquote {\bibinfo
  {title} {Nutrient transport driven by microbial active carpets},}\ }\href
  {\doibase 10.1103/PhysRevLett.121.248101} {\bibfield  {journal} {\bibinfo
  {journal} {Phys. Rev. Lett.}\ }\textbf {\bibinfo {volume} {121}},\ \bibinfo
  {pages} {248101} (\bibinfo {year} {2018})}\BibitemShut {NoStop}%
\bibitem [{\citenamefont {Jin}\ and\ \citenamefont
  {Riedel-Kruse}(2018)}]{jin2018biofilm}%
  \BibitemOpen
  \bibfield  {author} {\bibinfo {author} {\bibfnamefont {X.}~\bibnamefont
  {Jin}}\ and\ \bibinfo {author} {\bibfnamefont {I.~H.}\ \bibnamefont
  {Riedel-Kruse}},\ }\bibfield  {title} {\enquote {\bibinfo {title} {Biofilm
  lithography enables high-resolution cell patterning via optogenetic adhesin
  expression},}\ }\href {\doibase 10.1073/pnas.1720676115} {\bibfield
  {journal} {\bibinfo  {journal} {Proc. Nat. Acad. Sci.}\ }\textbf {\bibinfo
  {volume} {115}},\ \bibinfo {pages} {3698--3703} (\bibinfo {year}
  {2018})}\BibitemShut {NoStop}%
\bibitem [{\citenamefont {Schaller}\ \emph {et~al.}(2010)\citenamefont
  {Schaller}, \citenamefont {Weber}, \citenamefont {Semmrich}, \citenamefont
  {Frey},\ and\ \citenamefont {Bausch}}]{schaller2010polar}%
  \BibitemOpen
  \bibfield  {author} {\bibinfo {author} {\bibfnamefont {V.}~\bibnamefont
  {Schaller}}, \bibinfo {author} {\bibfnamefont {C.}~\bibnamefont {Weber}},
  \bibinfo {author} {\bibfnamefont {C.}~\bibnamefont {Semmrich}}, \bibinfo
  {author} {\bibfnamefont {E.}~\bibnamefont {Frey}}, \ and\ \bibinfo {author}
  {\bibfnamefont {A.~R.}\ \bibnamefont {Bausch}},\ }\bibfield  {title}
  {\enquote {\bibinfo {title} {Polar patterns of driven filaments},}\ }\href
  {\doibase 10.1038/nature09312} {\bibfield  {journal} {\bibinfo  {journal}
  {Nature}\ }\textbf {\bibinfo {volume} {467}},\ \bibinfo {pages} {73}
  (\bibinfo {year} {2010})}\BibitemShut {NoStop}%
\bibitem [{\citenamefont {Dittrich}\ and\ \citenamefont
  {Manz}(2006)}]{dittrich2006lab}%
  \BibitemOpen
  \bibfield  {author} {\bibinfo {author} {\bibfnamefont {P.~S.}\ \bibnamefont
  {Dittrich}}\ and\ \bibinfo {author} {\bibfnamefont {A.}~\bibnamefont
  {Manz}},\ }\bibfield  {title} {\enquote {\bibinfo {title} {Lab-on-a-chip:
  microfluidics in drug discovery},}\ }\href {\doibase 10.1038/nrd1985}
  {\bibfield  {journal} {\bibinfo  {journal} {Nat. Rev. Drug Discovery}\
  }\textbf {\bibinfo {volume} {5}},\ \bibinfo {pages} {210--218} (\bibinfo
  {year} {2006})}\BibitemShut {NoStop}%
\bibitem [{\citenamefont {Schneider}\ \emph {et~al.}(2011)\citenamefont
  {Schneider}, \citenamefont {Mandre},\ and\ \citenamefont
  {Brenner}}]{schneider2011algorithm}%
  \BibitemOpen
  \bibfield  {author} {\bibinfo {author} {\bibfnamefont {T.~M.}\ \bibnamefont
  {Schneider}}, \bibinfo {author} {\bibfnamefont {S.}~\bibnamefont {Mandre}}, \
  and\ \bibinfo {author} {\bibfnamefont {M.~P.}\ \bibnamefont {Brenner}},\
  }\bibfield  {title} {\enquote {\bibinfo {title} {Algorithm for a microfluidic
  assembly line},}\ }\href {\doibase 10.1103/PhysRevLett.106.094503} {\bibfield
   {journal} {\bibinfo  {journal} {Phys. Rev. Lett.}\ }\textbf {\bibinfo
  {volume} {106}},\ \bibinfo {pages} {094503} (\bibinfo {year}
  {2011})}\BibitemShut {NoStop}%
\bibitem [{\citenamefont {Nakamura}\ \emph {et~al.}(2014)\citenamefont
  {Nakamura}, \citenamefont {Chen}, \citenamefont {Howes}, \citenamefont
  {Schindler}, \citenamefont {Nogales},\ and\ \citenamefont
  {Bryant}}]{nakamura2014remote}%
  \BibitemOpen
  \bibfield  {author} {\bibinfo {author} {\bibfnamefont {M.}~\bibnamefont
  {Nakamura}}, \bibinfo {author} {\bibfnamefont {L.}~\bibnamefont {Chen}},
  \bibinfo {author} {\bibfnamefont {S.~C.}\ \bibnamefont {Howes}}, \bibinfo
  {author} {\bibfnamefont {T.~D.}\ \bibnamefont {Schindler}}, \bibinfo {author}
  {\bibfnamefont {E.}~\bibnamefont {Nogales}}, \ and\ \bibinfo {author}
  {\bibfnamefont {Z.}~\bibnamefont {Bryant}},\ }\bibfield  {title} {\enquote
  {\bibinfo {title} {Remote control of myosin and kinesin motors using
  light-activated gearshifting},}\ }\href {\doibase 10.1038/nnano.2014.147}
  {\bibfield  {journal} {\bibinfo  {journal} {Nat. Nanotech.}\ }\textbf
  {\bibinfo {volume} {9}},\ \bibinfo {pages} {693} (\bibinfo {year}
  {2014})}\BibitemShut {NoStop}%
\bibitem [{\citenamefont {Ruijgrok}\ \emph {et~al.}()\citenamefont {Ruijgrok},
  \citenamefont {Ghosh}, \citenamefont {Nakamura}, \citenamefont {Zemsky},
  \citenamefont {Chen}, \citenamefont {Vachharajani}, \citenamefont
  {Liphardt},\ and\ \citenamefont {Bryant}}]{ruijgrok2020newpaper}%
  \BibitemOpen
  \bibfield  {author} {\bibinfo {author} {\bibfnamefont {P.}~\bibnamefont
  {Ruijgrok}}, \bibinfo {author} {\bibfnamefont {R.~P.}\ \bibnamefont {Ghosh}},
  \bibinfo {author} {\bibfnamefont {M.}~\bibnamefont {Nakamura}}, \bibinfo
  {author} {\bibfnamefont {S.}~\bibnamefont {Zemsky}}, \bibinfo {author}
  {\bibfnamefont {R.}~\bibnamefont {Chen}}, \bibinfo {author} {\bibfnamefont
  {V.~T.}\ \bibnamefont {Vachharajani}}, \bibinfo {author} {\bibfnamefont
  {J.~T.}\ \bibnamefont {Liphardt}}, \ and\ \bibinfo {author} {\bibfnamefont
  {Z.}~\bibnamefont {Bryant}},\ }\bibfield  {title} {\enquote {\bibinfo {title}
  {Optical control of fast and processive engineered myosins in vitro and in
  living cells},}\ }\href@noop {} {\bibinfo  {journal} {Under review.}\
  }\BibitemShut {NoStop}%
\bibitem [{\citenamefont {Zhang}\ \emph {et~al.}(2019)\citenamefont {Zhang},
  \citenamefont {Redford}, \citenamefont {Ruijgrok}, \citenamefont {Kumar},
  \citenamefont {Mozaffari}, \citenamefont {Zemsky}, \citenamefont {Dinner},
  \citenamefont {Vitelli}, \citenamefont {Bryant}, \citenamefont {Gardel},\
  and\ \citenamefont {De~Pablo}}]{zhang2019structuring}%
  \BibitemOpen
\bibfield  {journal} {  }\bibfield  {author} {\bibinfo {author} {\bibfnamefont
  {R.}~\bibnamefont {Zhang}}, \bibinfo {author} {\bibfnamefont {S.~A.}\
  \bibnamefont {Redford}}, \bibinfo {author} {\bibfnamefont {P.~V.}\
  \bibnamefont {Ruijgrok}}, \bibinfo {author} {\bibfnamefont {N.}~\bibnamefont
  {Kumar}}, \bibinfo {author} {\bibfnamefont {A.}~\bibnamefont {Mozaffari}},
  \bibinfo {author} {\bibfnamefont {S.}~\bibnamefont {Zemsky}}, \bibinfo
  {author} {\bibfnamefont {A.~R.}\ \bibnamefont {Dinner}}, \bibinfo {author}
  {\bibfnamefont {V.}~\bibnamefont {Vitelli}}, \bibinfo {author} {\bibfnamefont
  {Z.}~\bibnamefont {Bryant}}, \bibinfo {author} {\bibfnamefont {M.~L.}\
  \bibnamefont {Gardel}}, \ and\ \bibinfo {author} {\bibfnamefont {J.~J.}\
  \bibnamefont {De~Pablo}},\ }\bibfield  {title} {\enquote {\bibinfo {title}
  {Structuring stress for active materials control},}\ }\href
  {https://arxiv.org/abs/1912.01630} {\bibfield  {journal} {\bibinfo  {journal}
  {arXiv preprint}\ ,\ \bibinfo {pages} {1912.01630}} (\bibinfo {year}
  {2019})}\BibitemShut {NoStop}%
\bibitem [{\citenamefont {Quinlan}(2016)}]{oocyte_streaming_review}%
  \BibitemOpen
  \bibfield  {author} {\bibinfo {author} {\bibfnamefont {M.~E.}\ \bibnamefont
  {Quinlan}},\ }\bibfield  {title} {\enquote {\bibinfo {title} {Cytoplasmic
  streaming in the drosophila oocyte},}\ }\href {\doibase
  10.1146/annurev-cellbio-111315-125416} {\bibfield  {journal} {\bibinfo
  {journal} {Ann. Rev. Cell Develop. Biol.}\ }\textbf {\bibinfo {volume}
  {32}},\ \bibinfo {pages} {173--195} (\bibinfo {year} {2016})},\ \bibinfo
  {note} {pMID: 27362645}\BibitemShut {NoStop}%
\bibitem [{\citenamefont {Monteith}\ \emph {et~al.}(2016)\citenamefont
  {Monteith}, \citenamefont {Brunner}, \citenamefont {Djagaeva}, \citenamefont
  {Bielecki}, \citenamefont {Deutsch},\ and\ \citenamefont
  {Saxton}}]{monteith2016mechanism}%
  \BibitemOpen
  \bibfield  {author} {\bibinfo {author} {\bibfnamefont {C.~E.}\ \bibnamefont
  {Monteith}}, \bibinfo {author} {\bibfnamefont {M.~E.}\ \bibnamefont
  {Brunner}}, \bibinfo {author} {\bibfnamefont {I.}~\bibnamefont {Djagaeva}},
  \bibinfo {author} {\bibfnamefont {A.~M.}\ \bibnamefont {Bielecki}}, \bibinfo
  {author} {\bibfnamefont {J.~M.}\ \bibnamefont {Deutsch}}, \ and\ \bibinfo
  {author} {\bibfnamefont {W.~M.}\ \bibnamefont {Saxton}},\ }\bibfield  {title}
  {\enquote {\bibinfo {title} {A mechanism for cytoplasmic streaming:
  Kinesin-driven alignment of microtubules and fast fluid flows},}\ }\href
  {\doibase 10.1016/j.bpj.2016.03.036} {\bibfield  {journal} {\bibinfo
  {journal} {Biophys. J.}\ }\textbf {\bibinfo {volume} {110}},\ \bibinfo
  {pages} {2053--2065} (\bibinfo {year} {2016})}\BibitemShut {NoStop}%
\bibitem [{\citenamefont {Osada}\ \emph {et~al.}(2016)\citenamefont {Osada},
  \citenamefont {Kawamura},\ and\ \citenamefont
  {Sano}}]{hydrogels_cytoskeleton}%
  \BibitemOpen
  \bibfield  {author} {\bibinfo {author} {\bibfnamefont {Y.}~\bibnamefont
  {Osada}}, \bibinfo {author} {\bibfnamefont {R.}~\bibnamefont {Kawamura}}, \
  and\ \bibinfo {author} {\bibfnamefont {K.}~\bibnamefont {Sano}},\ }\href@noop
  {} {\emph {\bibinfo {title} {Hydrogels of Cytoskeletal Proteins}}}\ (\bibinfo
   {publisher} {Springer},\ \bibinfo {year} {2016})\BibitemShut {NoStop}%
\bibitem [{\citenamefont {Sanchez}\ \emph {et~al.}(2012)\citenamefont
  {Sanchez}, \citenamefont {Chen}, \citenamefont {Decamp}, \citenamefont
  {Heymann},\ and\ \citenamefont {Dogic}}]{dogic2012}%
  \BibitemOpen
  \bibfield  {author} {\bibinfo {author} {\bibfnamefont {T.}~\bibnamefont
  {Sanchez}}, \bibinfo {author} {\bibfnamefont {D.}~\bibnamefont {Chen}},
  \bibinfo {author} {\bibfnamefont {S.}~\bibnamefont {Decamp}}, \bibinfo
  {author} {\bibfnamefont {M.}~\bibnamefont {Heymann}}, \ and\ \bibinfo
  {author} {\bibfnamefont {Z.}~\bibnamefont {Dogic}},\ }\bibfield  {title}
  {\enquote {\bibinfo {title} {{Spontaneous motion in hierarchically assembled
  active matter}},}\ }\href {\doibase 10.1038/nature11591} {\bibfield
  {journal} {\bibinfo  {journal} {Nature}\ }\textbf {\bibinfo {volume} {491}},\
  \bibinfo {pages} {431–434} (\bibinfo {year} {2012})}\BibitemShut {NoStop}%
\bibitem [{\citenamefont {Lighthill}(1952)}]{lighthill1952squirming}%
  \BibitemOpen
  \bibfield  {author} {\bibinfo {author} {\bibfnamefont {M.~J.}\ \bibnamefont
  {Lighthill}},\ }\bibfield  {title} {\enquote {\bibinfo {title} {On the
  squirming motion of nearly spherical deformable bodies through liquids at
  very small {R}eynolds numbers},}\ }\href {\doibase 10.1002/cpa.3160050201}
  {\bibfield  {journal} {\bibinfo  {journal} {Comm. Pure Appl. Math.}\ }\textbf
  {\bibinfo {volume} {5}},\ \bibinfo {pages} {109--118} (\bibinfo {year}
  {1952})}\BibitemShut {NoStop}%
\bibitem [{\citenamefont {Blake}(1971)}]{blake1971spherical}%
  \BibitemOpen
  \bibfield  {author} {\bibinfo {author} {\bibfnamefont {J.~R.}\ \bibnamefont
  {Blake}},\ }\bibfield  {title} {\enquote {\bibinfo {title} {A spherical
  envelope approach to ciliary propulsion},}\ }\href {\doibase
  10.1017/S002211207100048X} {\bibfield  {journal} {\bibinfo  {journal} {J.
  Fluid Mech.}\ }\textbf {\bibinfo {volume} {46}},\ \bibinfo {pages} {199--208}
  (\bibinfo {year} {1971})}\BibitemShut {NoStop}%
\bibitem [{\citenamefont {Reigh}\ \emph {et~al.}(2017)\citenamefont {Reigh},
  \citenamefont {Zhu}, \citenamefont {Gallaire},\ and\ \citenamefont
  {Lauga}}]{reigh2017swimming}%
  \BibitemOpen
  \bibfield  {author} {\bibinfo {author} {\bibfnamefont {S.~Y.}\ \bibnamefont
  {Reigh}}, \bibinfo {author} {\bibfnamefont {L.}~\bibnamefont {Zhu}}, \bibinfo
  {author} {\bibfnamefont {F.}~\bibnamefont {Gallaire}}, \ and\ \bibinfo
  {author} {\bibfnamefont {E.}~\bibnamefont {Lauga}},\ }\bibfield  {title}
  {\enquote {\bibinfo {title} {Swimming with a cage: low-reynolds-number
  locomotion inside a droplet},}\ }\href {\doibase 10.1039/C6SM01636G}
  {\bibfield  {journal} {\bibinfo  {journal} {Soft Matter}\ }\textbf {\bibinfo
  {volume} {13}},\ \bibinfo {pages} {3161--3173} (\bibinfo {year}
  {2017})}\BibitemShut {NoStop}%
\bibitem [{\citenamefont {Pedley}(2016)}]{pedley2016spherical}%
  \BibitemOpen
  \bibfield  {author} {\bibinfo {author} {\bibfnamefont {T.}~\bibnamefont
  {Pedley}},\ }\bibfield  {title} {\enquote {\bibinfo {title} {Spherical
  squirmers: models for swimming micro-organisms},}\ }\href {\doibase
  10.1093/imamat/hxw030} {\bibfield  {journal} {\bibinfo  {journal} {IMA J.
  Appl. Math.}\ }\textbf {\bibinfo {volume} {81}},\ \bibinfo {pages} {488--521}
  (\bibinfo {year} {2016})}\BibitemShut {NoStop}%
\bibitem [{\citenamefont {Pak}\ and\ \citenamefont
  {Lauga}(2014)}]{pak2014generalized}%
  \BibitemOpen
  \bibfield  {author} {\bibinfo {author} {\bibfnamefont {O.~S.}\ \bibnamefont
  {Pak}}\ and\ \bibinfo {author} {\bibfnamefont {E.}~\bibnamefont {Lauga}},\
  }\bibfield  {title} {\enquote {\bibinfo {title} {Generalized squirming motion
  of a sphere},}\ }\href {\doibase 10.1007/s10665-014-9690-9} {\bibfield
  {journal} {\bibinfo  {journal} {J. Eng. Math.}\ }\textbf {\bibinfo {volume}
  {88}},\ \bibinfo {pages} {1--28} (\bibinfo {year} {2014})}\BibitemShut
  {NoStop}%
\bibitem [{\citenamefont {Fadda}\ \emph {et~al.}(2020)\citenamefont {Fadda},
  \citenamefont {Molina},\ and\ \citenamefont {Yamamoto}}]{fadda2020dynamics}%
  \BibitemOpen
  \bibfield  {author} {\bibinfo {author} {\bibfnamefont {F.}~\bibnamefont
  {Fadda}}, \bibinfo {author} {\bibfnamefont {J.~J.}\ \bibnamefont {Molina}}, \
  and\ \bibinfo {author} {\bibfnamefont {R.}~\bibnamefont {Yamamoto}},\
  }\bibfield  {title} {\enquote {\bibinfo {title} {Dynamics of a chiral swimmer
  sedimenting on a flat plate},}\ }\href {https://arxiv.org/abs/2001.03326}
  {\bibfield  {journal} {\bibinfo  {journal} {arXiv preprint}\ ,\ \bibinfo
  {pages} {2001.03326}} (\bibinfo {year} {2020})}\BibitemShut {NoStop}%
\bibitem [{\citenamefont {Happel}\ and\ \citenamefont
  {Brenner}(1983)}]{happel2012low}%
  \BibitemOpen
  \bibfield  {author} {\bibinfo {author} {\bibfnamefont {J.}~\bibnamefont
  {Happel}}\ and\ \bibinfo {author} {\bibfnamefont {H.}~\bibnamefont
  {Brenner}},\ }\href {\doibase 10.1007/978-94-009-8352-6} {\emph {\bibinfo
  {title} {Low {R}eynolds number hydrodynamics}}}\ (\bibinfo  {publisher}
  {Springer Netherlands},\ \bibinfo {year} {1983})\BibitemShut {NoStop}%
\bibitem [{\citenamefont {Dauparas}\ and\ \citenamefont
  {Lauga}(2018)}]{LaugaStokesCorner}%
  \BibitemOpen
  \bibfield  {author} {\bibinfo {author} {\bibfnamefont {J.}~\bibnamefont
  {Dauparas}}\ and\ \bibinfo {author} {\bibfnamefont {E.}~\bibnamefont
  {Lauga}},\ }\bibfield  {title} {\enquote {\bibinfo {title} {Leading-order
  {S}tokes flows near a corner},}\ }\href {\doibase 10.1093/imamat/hxy014}
  {\bibfield  {journal} {\bibinfo  {journal} {IMA J. Appl. Math.}\ }\textbf
  {\bibinfo {volume} {83}},\ \bibinfo {pages} {590--633} (\bibinfo {year}
  {2018})}\BibitemShut {NoStop}%
\bibitem [{\citenamefont {Bajer}\ and\ \citenamefont
  {Moffatt}(1990)}]{bajer1990class}%
  \BibitemOpen
  \bibfield  {author} {\bibinfo {author} {\bibfnamefont {K.}~\bibnamefont
  {Bajer}}\ and\ \bibinfo {author} {\bibfnamefont {H.}~\bibnamefont
  {Moffatt}},\ }\bibfield  {title} {\enquote {\bibinfo {title} {On a class of
  steady confined stokes flows with chaotic streamlines},}\ }\href {\doibase
  10.1017/S0022112090001999} {\bibfield  {journal} {\bibinfo  {journal} {J.
  Fluid Mech.}\ }\textbf {\bibinfo {volume} {212}},\ \bibinfo {pages}
  {337--363} (\bibinfo {year} {1990})}\BibitemShut {NoStop}%
\bibitem [{\citenamefont {Stone}\ \emph {et~al.}(1991)\citenamefont {Stone},
  \citenamefont {Nadim},\ and\ \citenamefont {Strogatz}}]{stone1991chaotic}%
  \BibitemOpen
  \bibfield  {author} {\bibinfo {author} {\bibfnamefont {H.~A.}\ \bibnamefont
  {Stone}}, \bibinfo {author} {\bibfnamefont {A.}~\bibnamefont {Nadim}}, \ and\
  \bibinfo {author} {\bibfnamefont {S.~H.}\ \bibnamefont {Strogatz}},\
  }\bibfield  {title} {\enquote {\bibinfo {title} {Chaotic streamlines inside
  drops immersed in steady stokes flows},}\ }\href {\doibase
  10.1017/S002211209100383X} {\bibfield  {journal} {\bibinfo  {journal} {J.
  Fluid Mech.}\ }\textbf {\bibinfo {volume} {232}},\ \bibinfo {pages}
  {629--646} (\bibinfo {year} {1991})}\BibitemShut {NoStop}%
\bibitem [{\citenamefont {Ward}\ and\ \citenamefont
  {Homsy}(2006)}]{ward2006chaotic}%
  \BibitemOpen
  \bibfield  {author} {\bibinfo {author} {\bibfnamefont {T.}~\bibnamefont
  {Ward}}\ and\ \bibinfo {author} {\bibfnamefont {G.}~\bibnamefont {Homsy}},\
  }\bibfield  {title} {\enquote {\bibinfo {title} {Chaotic streamlines in a
  translating drop with a uniform electric field},}\ }\href {\doibase
  10.1017/S0022112005007354} {\bibfield  {journal} {\bibinfo  {journal} {J.
  Fluid Mech.}\ }\textbf {\bibinfo {volume} {547}},\ \bibinfo {pages}
  {215--230} (\bibinfo {year} {2006})}\BibitemShut {NoStop}%
\bibitem [{\citenamefont {Smith}\ \emph {et~al.}(2019)\citenamefont {Smith},
  \citenamefont {Umbanhowar}, \citenamefont {Lueptow},\ and\ \citenamefont
  {Ottino}}]{smith2019geometry}%
  \BibitemOpen
  \bibfield  {author} {\bibinfo {author} {\bibfnamefont {L.~D.}\ \bibnamefont
  {Smith}}, \bibinfo {author} {\bibfnamefont {P.~B.}\ \bibnamefont
  {Umbanhowar}}, \bibinfo {author} {\bibfnamefont {R.~M.}\ \bibnamefont
  {Lueptow}}, \ and\ \bibinfo {author} {\bibfnamefont {J.~M.}\ \bibnamefont
  {Ottino}},\ }\bibfield  {title} {\enquote {\bibinfo {title} {The geometry of
  cutting and shuffling: An outline of possibilities for piecewise
  isometries},}\ }\href {\doibase 10.1016/j.physrep.2019.01.003} {\bibfield
  {journal} {\bibinfo  {journal} {Physics Reports}\ } (\bibinfo {year}
  {2019}),\ 10.1016/j.physrep.2019.01.003}\BibitemShut {NoStop}%
\bibitem [{\citenamefont {Ottino}\ and\ \citenamefont
  {Ottino}(1989)}]{ottino1989kinematics}%
  \BibitemOpen
  \bibfield  {author} {\bibinfo {author} {\bibfnamefont {J.~M.}\ \bibnamefont
  {Ottino}}\ and\ \bibinfo {author} {\bibfnamefont {J.}~\bibnamefont
  {Ottino}},\ }\href@noop {} {\emph {\bibinfo {title} {The kinematics of
  mixing: stretching, chaos, and transport}}},\ Vol.~\bibinfo {volume} {3}\
  (\bibinfo  {publisher} {Cambridge university press},\ \bibinfo {year}
  {1989})\BibitemShut {NoStop}%
\bibitem [{\citenamefont {Warner}\ and\ \citenamefont
  {Terentjev}(2007)}]{warner2007liquid}%
  \BibitemOpen
  \bibfield  {author} {\bibinfo {author} {\bibfnamefont {M.}~\bibnamefont
  {Warner}}\ and\ \bibinfo {author} {\bibfnamefont {E.~M.}\ \bibnamefont
  {Terentjev}},\ }\href
  {https://global.oup.com/academic/product/liquid-crystal-elastomers-9780198527671}
  {\emph {\bibinfo {title} {Liquid crystal elastomers}}}\ (\bibinfo
  {publisher} {Oxford University Press},\ \bibinfo {year} {2007})\BibitemShut
  {NoStop}%
\bibitem [{\citenamefont {Stuart}\ \emph {et~al.}(2010)\citenamefont {Stuart},
  \citenamefont {Huck}, \citenamefont {Genzer}, \citenamefont {M{\"u}ller},
  \citenamefont {Ober}, \citenamefont {Stamm}, \citenamefont {Sukhorukov},
  \citenamefont {Szleifer}, \citenamefont {Tsukruk}, \citenamefont {Urban}
  \emph {et~al.}}]{stuart2010emerging}%
  \BibitemOpen
  \bibfield  {author} {\bibinfo {author} {\bibfnamefont {M.~A.~C.}\
  \bibnamefont {Stuart}}, \bibinfo {author} {\bibfnamefont {W.~T.}\
  \bibnamefont {Huck}}, \bibinfo {author} {\bibfnamefont {J.}~\bibnamefont
  {Genzer}}, \bibinfo {author} {\bibfnamefont {M.}~\bibnamefont {M{\"u}ller}},
  \bibinfo {author} {\bibfnamefont {C.}~\bibnamefont {Ober}}, \bibinfo {author}
  {\bibfnamefont {M.}~\bibnamefont {Stamm}}, \bibinfo {author} {\bibfnamefont
  {G.~B.}\ \bibnamefont {Sukhorukov}}, \bibinfo {author} {\bibfnamefont
  {I.}~\bibnamefont {Szleifer}}, \bibinfo {author} {\bibfnamefont {V.~V.}\
  \bibnamefont {Tsukruk}}, \bibinfo {author} {\bibfnamefont {M.}~\bibnamefont
  {Urban}},  \emph {et~al.},\ }\bibfield  {title} {\enquote {\bibinfo {title}
  {Emerging applications of stimuli-responsive polymer materials},}\ }\href
  {\doibase 10.1038/nmat2614} {\bibfield  {journal} {\bibinfo  {journal} {Nat.
  Mater.}\ }\textbf {\bibinfo {volume} {9}},\ \bibinfo {pages} {101} (\bibinfo
  {year} {2010})}\BibitemShut {NoStop}%
\bibitem [{\citenamefont {Wang}\ \emph {et~al.}(2013)\citenamefont {Wang},
  \citenamefont {Desai},\ and\ \citenamefont {Lee}}]{wang2013light}%
  \BibitemOpen
  \bibfield  {author} {\bibinfo {author} {\bibfnamefont {E.}~\bibnamefont
  {Wang}}, \bibinfo {author} {\bibfnamefont {M.~S.}\ \bibnamefont {Desai}}, \
  and\ \bibinfo {author} {\bibfnamefont {S.-W.}\ \bibnamefont {Lee}},\
  }\bibfield  {title} {\enquote {\bibinfo {title} {Light-controlled
  graphene-elastin composite hydrogel actuators},}\ }\href {\doibase
  10.1021/nl401088b} {\bibfield  {journal} {\bibinfo  {journal} {Nano Lett.}\
  }\textbf {\bibinfo {volume} {13}},\ \bibinfo {pages} {2826--2830} (\bibinfo
  {year} {2013})}\BibitemShut {NoStop}%
\end{thebibliography}%

\newpage

\begin{figure*}
\includegraphics[width=\textwidth]{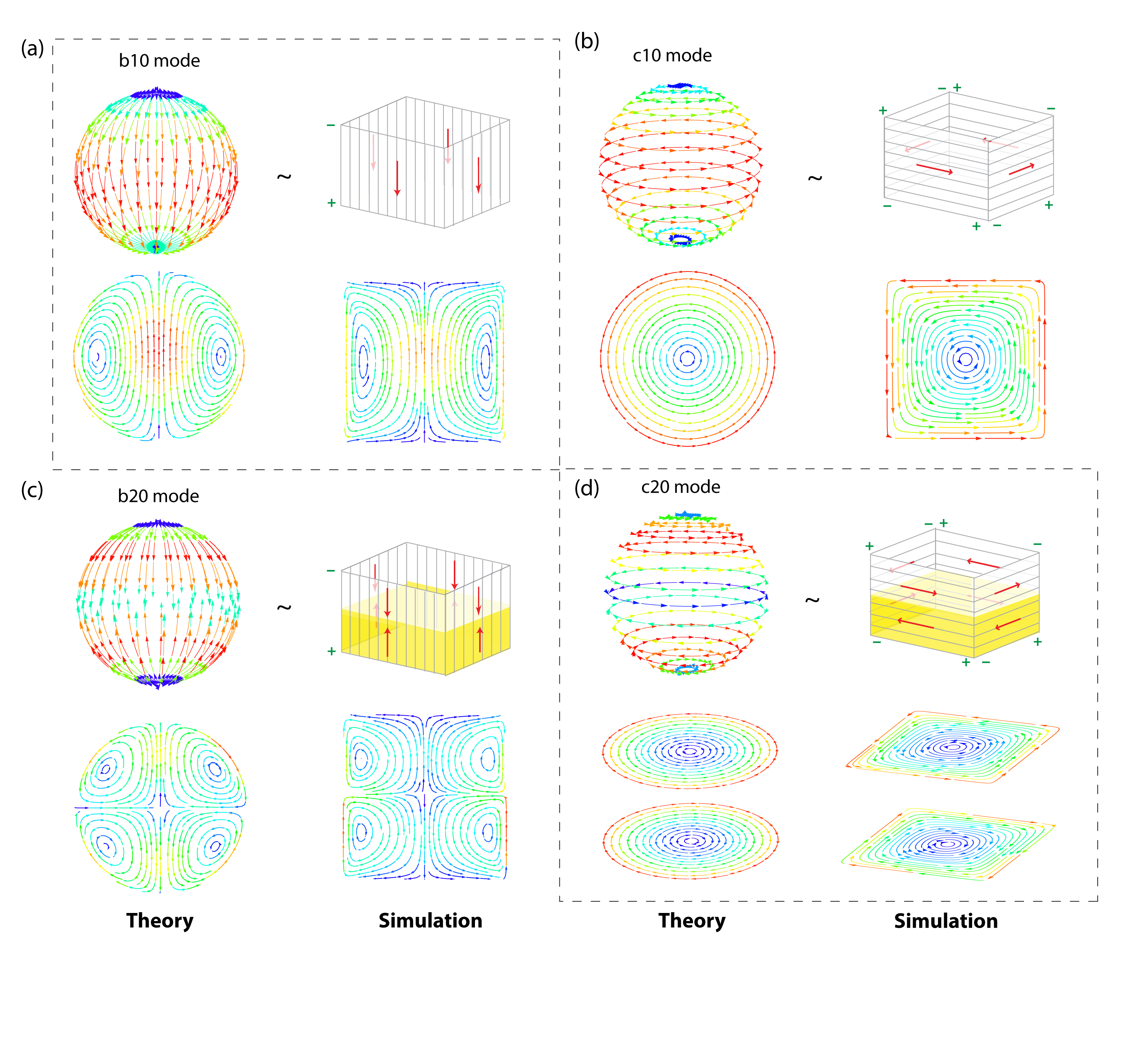}
\caption{\label{Fig4} 
Fundamental Modes of the Squirmer Model: an analytical approach to boundary driven flows. 
Shown in panels (a-d) are comparisons between theory (left) and simulations (right) for the first four axisymmetric modes. The green $\pm$ symbols at the corners of the box indicate the polarity of the actin filaments patterned on the boundary. The top row of each panel depicts the surface patterning, and the bottom row depicts the internal flow structures taken at some cross section. 
\textbf{(a,c)} The surface patterns of the $b_{10}$ and $b_{20}$ modes lie on longitudinal tracks. The $b_{10}$ mode consists of uniform (in direction) motion from the north to the south pole, while the $b_{20}$ mode naturally encodes a line of head-on defects at the equator. All interior flow structures are taken at the cross section $y=0$.
\textbf{(b,d)} Conversely, the surface patterns of the $c_{10}$ and $c_{20}$ modes lie on lines of latitude. Similar to the $b_{20}$, the $c_{20}$ mode naturally encodes a shear defect along its equator. The interior cross-sections of $c_{10}$ for both the sphere and the box are taken at $x=0$, while the pair of cross sections of $c_{20}$ are taken at $\theta = \pi/3, 3\pi/4$ for the sphere and at $z = 0.25, 0.75$.
Cross sections of the sphere and box in all four panels show that the interior flow structures are analogous. This shows that the language of head-on and shear defects are intrinsically built into the squirmer model. 
}
\end{figure*}

\newpage

\newpage

\begin{figure*}
\includegraphics[width=\textwidth]{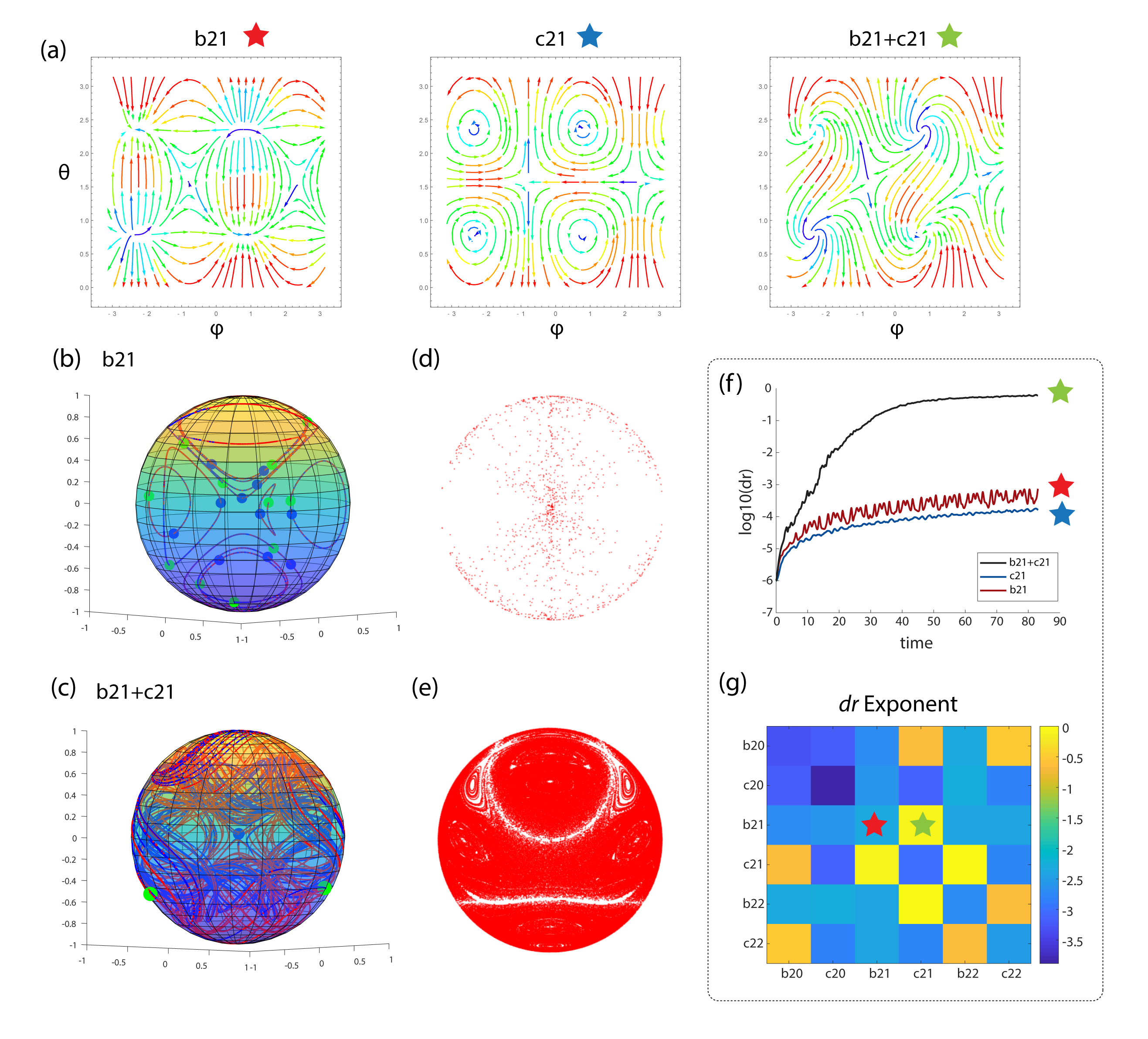}
\caption{\label{Fig5} 
Emergence of chaotic mixing. 
\textbf{(a)}
The surface streamline structure of the modes (i.e. the fluid velocity at r=1) is plotted as a function of the spherical angles $\theta$ and $\phi$. Note that the $b$ modes give rise to patches of oppositely moving flow, whereas the $c$ modes comprise of closed-vortices. %The superposition of the $b$ and $c$ modes adds a ``twist'' to the $b$-mode. 
\textbf{(b)} Examples of 10 separate trajectories of the $b_{21}$ mode. The blue dots denote the starting points of two particles, spaced $10^{-6}$ apart, a distance that cannot be resolved. The green dots denote the ending positions after integrating for $t=500$, which are still very closely spaced, showing that the two trajectories do not substantially diverge. 
\textbf{(c)} Example of a single chaotic trajectory of the $b_{21}+c_{21}$ mode. Note the positions of the green dots, which now span a distance comparable to the size of the droplet.
\textbf{(d)} The Poincare section at $x=0$ for the $b_{21}$ mode computed from 1000 trajectories. Note that the plane is only sparsely populated. 
\textbf{(e)} The Poincare section of $b_{21}+c_{21}$, on the other hand, nearly fills the entire plane. 
\textbf{(f)} The evolution in time of the logarithmic displacement between pairs of particles initially spaced $10^{-6}$ units apart, averaged over 1000 randomly seeded trajectories. These plots show that of the 3 modes in (a), only the superposed mode shows evidence of chaotic mixing on the scale of the droplet size. 
\textbf{(g)} Exponents of the time evolution of $dr$ after integrating for $t=500$ for various configurations of modes, indicated by the row $r$ and column $c$ of the element in the matrix. $(r,c) = (2,3)$ denotes the $c_{21} + b_{22}$ mode, for example. The matrix shows that the only modes that show evidence of chaotic mixing are the $b$ and $c$ superposed modes. However, not all such modes are chaotic, as evidenced by the $b_{21}+c_{22}$ mode.}
\end{figure*}

\newpage

\end{document}